\documentclass[fleqn,usenatbib]{mnras}

\usepackage[T1]{fontenc}
\usepackage{ae,aecompl}
 
\usepackage{graphicx}	
\usepackage{amsmath}	

\usepackage{amssymb}	
\usepackage{float}
\usepackage{caption}
\usepackage{subcaption}

\title[Radial profiles of the Ophiuchus Cluster. ]{ Chemical enrichment of ICM within the Ophiuchus cluster I: radial profiles.}

\author[Gatuzz et al.]{
Efrain Gatuzz$^{1}$\thanks{E-mail: egatuzz@mpe.mpg.de},
J. S. Sanders$^{1}$,
K. Dennerl$^{1}$,
A. Liu$^{1}$,
A. C. Fabian$^{2}$,
C. Pinto$^{3}$,\newauthor
D. Eckert$^{4}$ 
S. A. Walker$^{5}$
and J. ZuHone$^{6}$
\\
$^{1}$ Max-Planck-Institut f\"ur extraterrestrische Physik, Gie{\ss}enbachstra{\ss}e 1, 85748 Garching, Germany\\
$^{2}$ Institute of Astronomy, Madingley Road, Cambridge CB3 0HA, UK\\ 
$^{3}$ INAF - IASF Palermo, Via U. La Malfa 153, I-90146 Palermo, Italy \\
$^{4}$ Department of Astronomy, University of Geneva, Ch. d\rq Ecogia 16, CH-1290 Versoix, Switzerland \\ 
$^{5}$ Department of Physics and Astronomy, University of Alabama in Huntsville, Huntsville, AL 35899, USA\\
$^{6}$ Harvard-Smithsonian Center for Astrophysics, 60 Garden Street, Cambridge, MA, 02138, USA
}

\date{Accepted XXX. Received YYY; in original form ZZZ} 
\pubyear{2023} 
\begin{document}
 \label{firstpage}
\pagerange{\pageref{firstpage}--\pageref{lastpage}}
\maketitle 

\begin{abstract} 
The analysis of the elemental abundances in galaxy clusters offers valuable insights into the formation and evolution of galaxies.  In this study, we explore the chemical enrichment of the intergalactic medium (ICM) in the Ophiuchus cluster by utilizing {\it XMM-Newton} EPIC-pn observations. We explore the radial profiles of Si, S, Ar, Ca, and Fe. Due to the high absorption of the system, we have obtained only upper limits for O, Ne, Mg, and Ni. We model the X/Fe ratio profiles with a linear combination of core-collapse supernovae (SNcc) and type~Ia supernovae (SNIa) models. We found a flat radial distribution of SNIa ratio over the total cluster enrichment $10-30\%$ for all radii. However, the absence of light $\alpha$-elements abundances may lead to over-estimation of the SNcc contribution. 

\end{abstract} 
 


\section{Introduction}\label{sec_in}     
The distribution of element abundance within the diffuse intracluster medium (ICM) can be studied effectively using X-ray spectroscopy to study the intensity of emission lines \citep[e.g.,][for a recent review]{mer18}. Such studies have shown that many cool-core clusters exhibit an excess of Fe abundance, as well as other elements, in their central regions \citep[e.g.][]{deg01,chu03,pan15,mer17,liu19,liu20}, while outside of these regions, flat and azimuthally uniform Fe distributions have been observed in the outskirts \citep{mat11,wer13,sim15,urb17}. Observations of the Perseus cluster with {\it Hitomi} suggest that the abundance ratios near the cluster core are consistent with solar levels \citep{hit18,sim19}.  However, the abundances cannot be explained by simple combinations of SNcc and Ia, and including neutrino physics in the core-collapse supernova yield calculations may improve agreement with observed patterns of $\alpha$-elements in the Perseus Cluster core \citep[see e.g.][]{sim19}.

The physical properties and chemical composition of the ICM offer valuable information about the distribution and origin of chemical elements during the evolution of the Universe. SNcc are the primary source of light $\alpha$-elements (O, Ne, Mg), while SNIa mainly produce Fe-peak elements (Cr, Mn, Fe, Ni).  The synthesis of intermediate-mass elements (e.g., Si, S, Ar, and Ca) is done through the contribution of both SNcc and SNIa \citep[e.g.,][and references therein]{nom13}. Hence, the ICM metal content is sensitive to the number of SNIa and SNcc that contribute to chemical enrichment, the initial metallicity of the progenitors, the initial mass function (IMF) of the stars that explode as SNcc, the time scale over which the supernova products are expelled, and the SNIa explosion mechanism \citep[see][for a review]{wer08}. The ICM is infused with these elements from their host galaxies through galactic winds, AGN bubbles uplift and ram-pressure stripping.

Located at z = 0.0296 \citep{dur15}, the Ophiuchus cluster is a promising target to study the chemical enrichment of ICM due to its brightness in the 2--10 keV sky, ranking as the second brightest galaxy cluster and possessing the second strongest iron line after Perseus \citep{edg90,nev09}. 
Analyzing the only {\it XMM-Newton} observation centered in the cluster along with {\it INTEGRAL} data, \citet{nev09} determined that the non-thermal electron pressure is approximately 1$\%$ of the thermal electron pressure. 
A two-temperature model is necessary to fit the X-ray emission in the core, indicating a cooling core. The presence of cold fronts suggests significant sloshing of the X-ray emitting gas \citep{per09}. 
The truncated cool core is sharply peaked, with a temperature increase from kT $\sim$ 1 keV at the center to kT $\sim$ 9 keV at r$\sim$ 30 kpc. It is dynamically disturbed by Kelvin-Helmholtz and Rayleigh-Taylor instabilities \citep{mil10,wer16b}. 
\citet{liu19} found a drop in the Fe abundance near the cluster core. 
The AGN displays weak, point-like radio emission. However, \citet{gia20} suggests the presence of a large cavity southeast of the cluster, potentially a fossil of the most potent AGN outburst in any galaxy cluster. 
While massive and relatively relaxed, there may be a minor merger occurring in Ophiuchus \citep[e.g.][]{dur15}. 
\citet{gat23a} studied the velocity structure of the ICM within the Ophiuchus cluster. 
Their analysis shows that the velocities in the central regions are similar to the velocity of the brightest cluster galaxy. 
They also found a large interface region following a sharp surface brightness discontinuity at $250$~kpc from the cluster center, where the velocity changes abruptly from blueshifted to redshifted gas.

This paper is structured as follows. Section~\ref{sec_dat} provides a detailed description of the data reduction process. Section~\ref{sec_fits} discusses the fitting procedure used in our analysis. We then proceed to discuss these results in Section~\ref{sec_dis}. Finally, Section~\ref{sec_con} summarizes our findings and presents our conclusions. Throughout this paper, we assume the distance of Ophiuchus to be $z=0.0296$ \citep{dur15} and a concordance $\Lambda$CDM cosmology with $\Omega_m = 0.3$, $\Omega_\Lambda = 0.7$, and $H_{0} = 70 \textrm{ km s}^{-1}\ \textrm{Mpc}^{-1} $.

\section{Data reduction}\label{sec_dat}
Table~\ref{tab_obsids} shows the {\it XMM-Newton} observations analyzed, including IDs, coordinates, dates, and clean exposure times. The observations are the same as used in \citet{gat23a} and we followed the same data reduction process. We reduced the {\it XMM-Newton} European Photon Imaging Camera \citep[EPIC,][]{str01} spectra with the Science Analysis System (SAS\footnote{\url{https://www.cosmos.esa.int/web/xmm-newton/sas}}, version 19.1.0). We processed each observation with the {\tt epchain} SAS tool, using only single-pixel events (PATTERN==0) and filtering bad time intervals from flares by applying a 1.0 cts/s rate threshold. We also use FLAG==0 as a filtering parameter in order to avoid regions close to CCD edges and bad pixels.

In order to measure the ICM velocity structure, we follow the method described in \citet[][c]{san20,gat22a,gat22b,gat23a}. We created the final event files for each observation using a new energy calibration scale which allow us to measure velocities with uncertainties down to 150 km/s at the Fe-K complex. This is done by using the background X-ray instrumental lines identified in the detector spectra as references for the absolute energy scale. Point sources were identified using the SAS task {\tt edetect\_chain}, with a likelihood parameter {\tt det\_ml} $> 10$. These point sources were excluded from the subsequent analysis.

\section{Spectral fitting}\label{sec_fits}

\begin{figure}    
\centering
\includegraphics[width=0.45\textwidth]{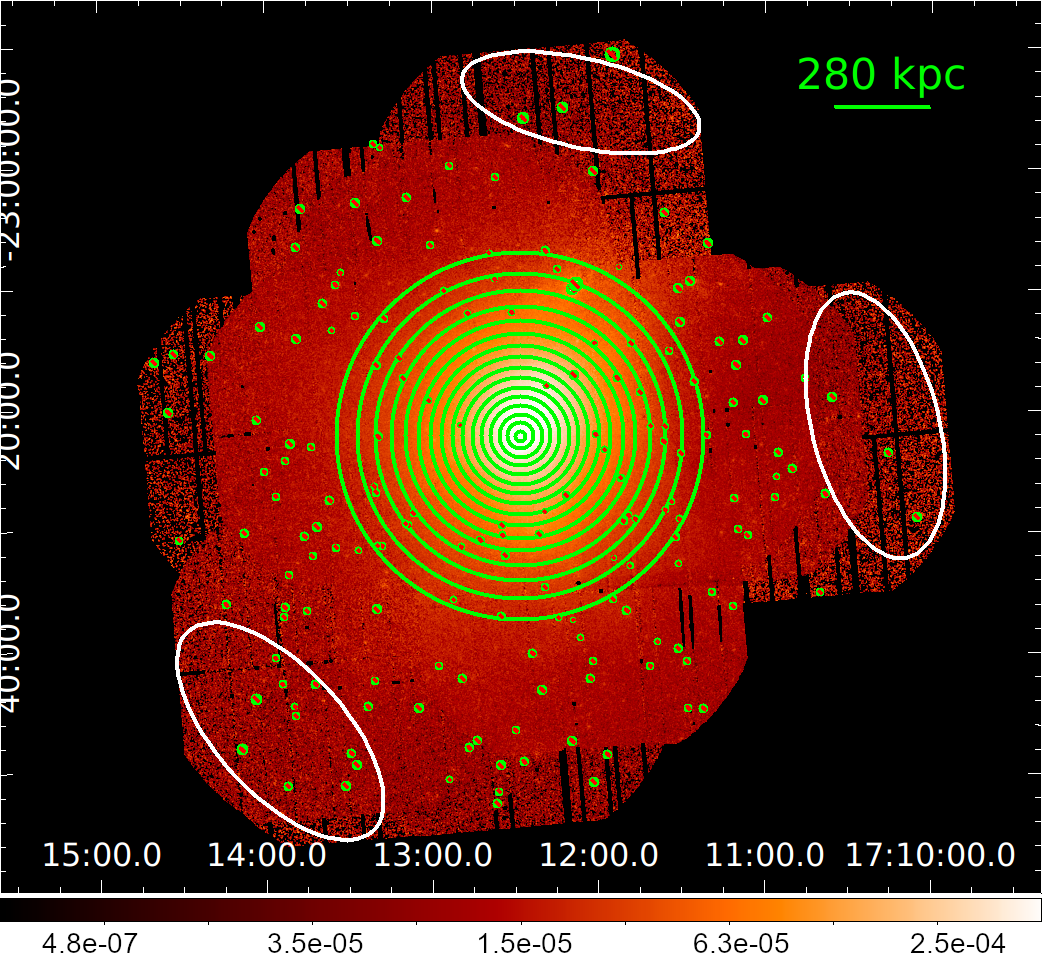} 
\caption{
Ophiuchus regions for which EPIC-pn spectra were extracted. The small circles correspond to point like sources excluded from the analysis. The white region was used to model the astrophysical background.
} \label{fig_regions} 
\end{figure}

In order to study the radial distribution of physical parameters in the ICM we created non-overlapping circular spectra extraction regions. The thickness of these rings increases as the square root distance from the Ophiuchus center. Figure~\ref{fig_regions} shows these regions.  We combined the spectra from different observations for each spatial region. The new EPIC-pn energy calibration scale cannot be applied for lower energies. Therefore, we load the data twice to fit separately but simultaneously the soft (0.5-4~keV) and hard (4-10~keV) energy bands. Then, we only set the redshift as a free parameter for the 4-10~keV energy band. By including the low energy band, we have better constrain for temperatures and metallicities. We analyzed the spectra with the {\it xspec} spectral fitting package (version 12.13.0\footnote{\url{https://heasarc.gsfc.nasa.gov/xanadu/xspec/}}). We assumed {\tt cash} statistics \citep{cas79}. Errors are quoted at 1$\sigma$ confidence level unless otherwise stated. Abundances are given relative to \citet{lod09}.

It has been shown that the Ophiuchus cluster requires a multi-temperature component to model the central emission \citep{nev09,wer16}. Therefore, we model the ICM emission spectra with the {\tt lognorm} model, which considers a log-normal temperature distribution \citep{fra13,gat22b}. The parameters of the model are the width of the temperature distribution in log space ($\sigma$), central temperature ($T$), abundances for individual elements (O, Ne, Si, Ar, S, Ca, Fe, Ni), redshift and normalization. The column density is also set as a free parameter (more details in Section~\ref{sec_nh}). As discussed in \citet{gat23b}, the Mg and Al abundances may not be reliable due to a strong Al K$\alpha$ instrumental line $\sim 1.5$ keV. It has been shown that, when fitting the EPIC-pn spectra over a large range, the continuum emission may be slightly under or overestimated in some specific bands \citep[see a detailed discussion in ][]{mer15,mer16}. In order to minimize such biases, we re-fit the spectra locally by fixing the temperature parameters (i.e., $T$, $\log\sigma$) to the values obtained from the global fit. The energy ranges for the local fits are $0.50-0.80$ keV (for O), $0.90-1.20$ keV (for Ne), $1.65-2.26$ keV (for Si), $2.27-3$ keV (for S and Ar), $3.5-4.5$ keV (for Ca), $6.25-7.25$ keV (for Fe) and $7.24-8.02$ keV (for Ni). Hereafter, all the best-fit abundances are locally corrected.

We consider the instrumental as well as the astrophysical background as follows. We included instrumental emission lines due to  Cu-$K\alpha$, Cu-$K\beta$, Ni-$K\alpha$, Zn-$K\alpha$ and Al-$\alpha$ transitions in combination with a power-law component with a photon index fixed to 0.136 \citep[i.e., the average value obtained from the archival observations analyzed in ][]{san20}. For the astrophysical background, we consider an unabsorbed thermal plasma model ({\tt apec}) for the Local Hot Bubble (LHB), one absorbed thermal plasma model for the Galactic halo (GH) emission, and a power-law with $\Gamma=1.45$ to account for the unresolved population of point sources \citep[e.g.][]{yos09}. We estimated temperatures of these astrophysical background components by modeling the spectrum from 3 regions located in the outskirts of the cluster (see Figure~\ref{fig_regions}). The best-fit parameters obtained for the LHB, GH, and CXB components are listed in Table~\ref{tab_bkg}. Thus, we fixed the temperatures of the background components to the best-fit values of $T_{e}=0.738$~keV and $T_{e}=0.202$~keV, while the normalization is a free parameter in the model. While we have identified these components as part of the LHB and GH, such temperatures could trace the virial temperature in the Milky Way \citep[$\sim$0.2~keV, see ][]{das21} and the eROSITA bubbles \citep[$\sim$0.74~keV, see ][]{gup23}, given the location of the Ophiuchus cluster near the Galactic plane.

  \begin{table} 
\footnotesize
\caption{\label{tab_obsids}{\it XMM-Newton} observations of the Ophiuchus cluster.}
\centering 
\begin{tabular}{cccccccc}   
\hline
ObsID & RA & DEC & Date & Exposure \\
 &&&Start-time& (ks)\\
\hline
0105460101&17:12:36.00& -24:14:41.0&	2000-09-18&        21.3\\
0105460601&17:12:36.00& -24:14:41.0&	2001-08-30&        18.8\\
0206990701&17:09:44.89& -23:46:58.0&	2005-02-28&        18.3\\
0505150101&17:12:27.39& -23:22:19.7&	2007-09-02&        36.7\\
0505150201&17:12:27.39& -23:42:19.7&	2007-09-26&        26.7\\
0505150301&17:11:00.23& -23:22:18.1&	2008-02-20&        27.9\\
0505150401&17:12:27.39& -23:02:19.7&	2008-02-21&        27.9\\
0505150501&17:13:54.55& -23:22:18.3&	2008-02-21&        29.8\\
0862220501&17:13:14.99& -23:43:15.0&	2020-09-01&        44.9\\ 
0862220101&17:13:25.49& -23:41:15.0&	2020-08-31&        40.0\\
0862220301&17:13:32.69& -24:02:51.0&	2020-09-28&        38.5\\
0862220201&17:14:52.70& -23:43:08.0&	2021-03-18&        33.0\\ 
0880280801&17:11:57.84& -23:35:41.0&	2021-09-06&        59.9\\
0880280401&17:11:57.84& -23:35:41.0&	2021-09-17&       126.2\\
0880280901&17:11:57.84& -23:35:41.0&	2021-09-17&        118.7\\
0880280201&17:11:31.97& -23:21:05.1&	2021-09-19&       126.1\\
0880281001&17:11:31.97& -23:21:05.1&	2021-09-19&        119.8\\
0880280301&17:13:12.01& -23:21:35.4&	2021-09-21&       127.0\\
0880281101&17:13:12.01& -23:21:35.4&	2021-09-21&        11.0\\
0880280101&17:12:57.59& -23:11:15.8&	2021-09-29&       127.2\\
0880281201&17:12:57.59& -23:11:15.8&	2021-09-29&        11.0\\
0880280701&17:13:12.01& -23:21:35.4&	2022-02-26&        55.1\\
0880280601&17:11:31.97& -23:21:05.1&	2022-03-12&        54.0\\
0880281301&17:11:31.97& -23:21:05.1&	2022-03-11&         5.9\\
0880280501&17:12:57.59& -23:11:15.8&	2022-03-16&        51.0\\ 
 \\
 \hline
\end{tabular}
\end{table}

\begin{table}
\caption{Background model best-fit parameters. }\label{tab_bkg}
\begin{center}
\begin{tabular}{ccccccc}
\hline
\hline
  & $\Gamma$ & & kT (keV)     \\
\hline  
CXB & 1.45  & GH & $ 0.738 \pm 0.05 $   \\
 &  &  LHB & $0.202\pm 0.02$     \\
\hline 
\end{tabular}
\end{center}
  
\end{table}

\begin{figure*}    
\centering
\includegraphics[width=0.33\textwidth]{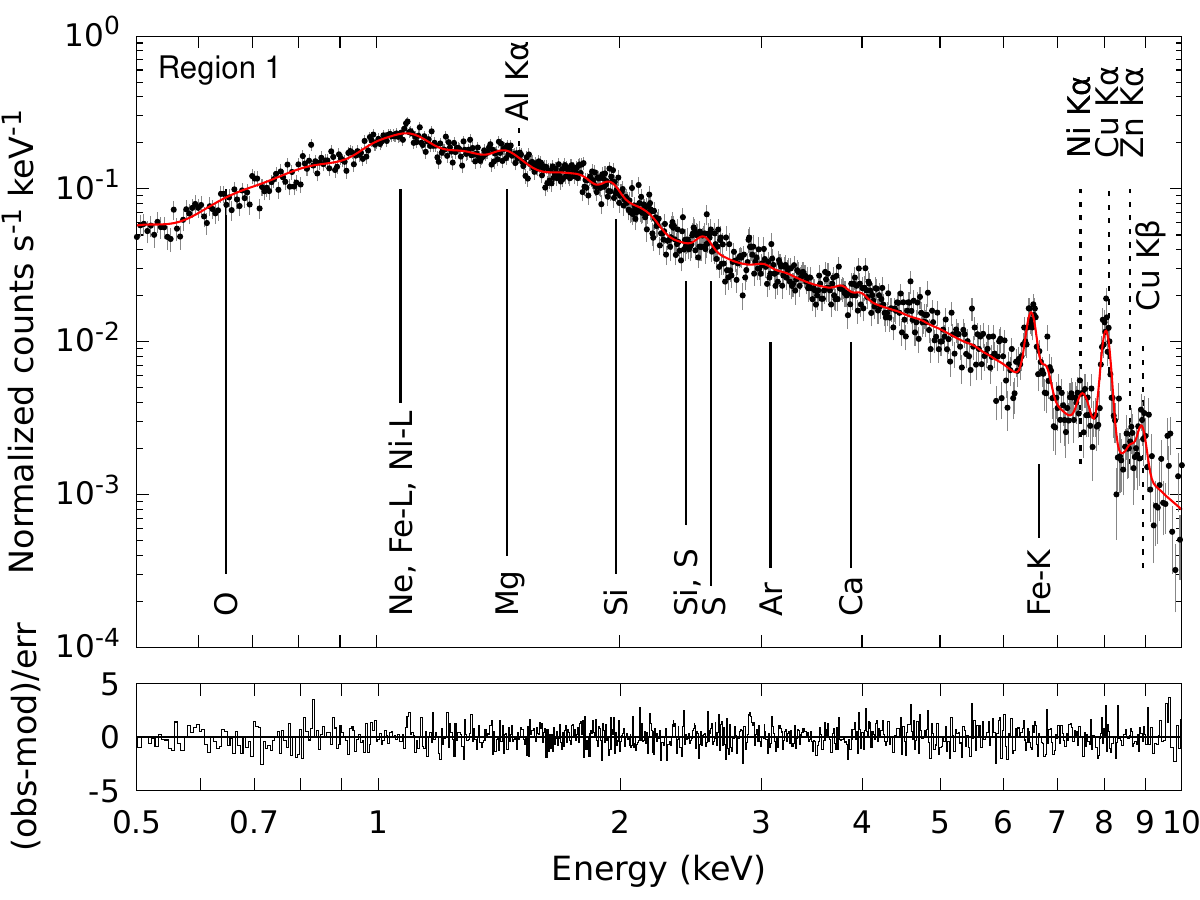} 
\includegraphics[width=0.33\textwidth]{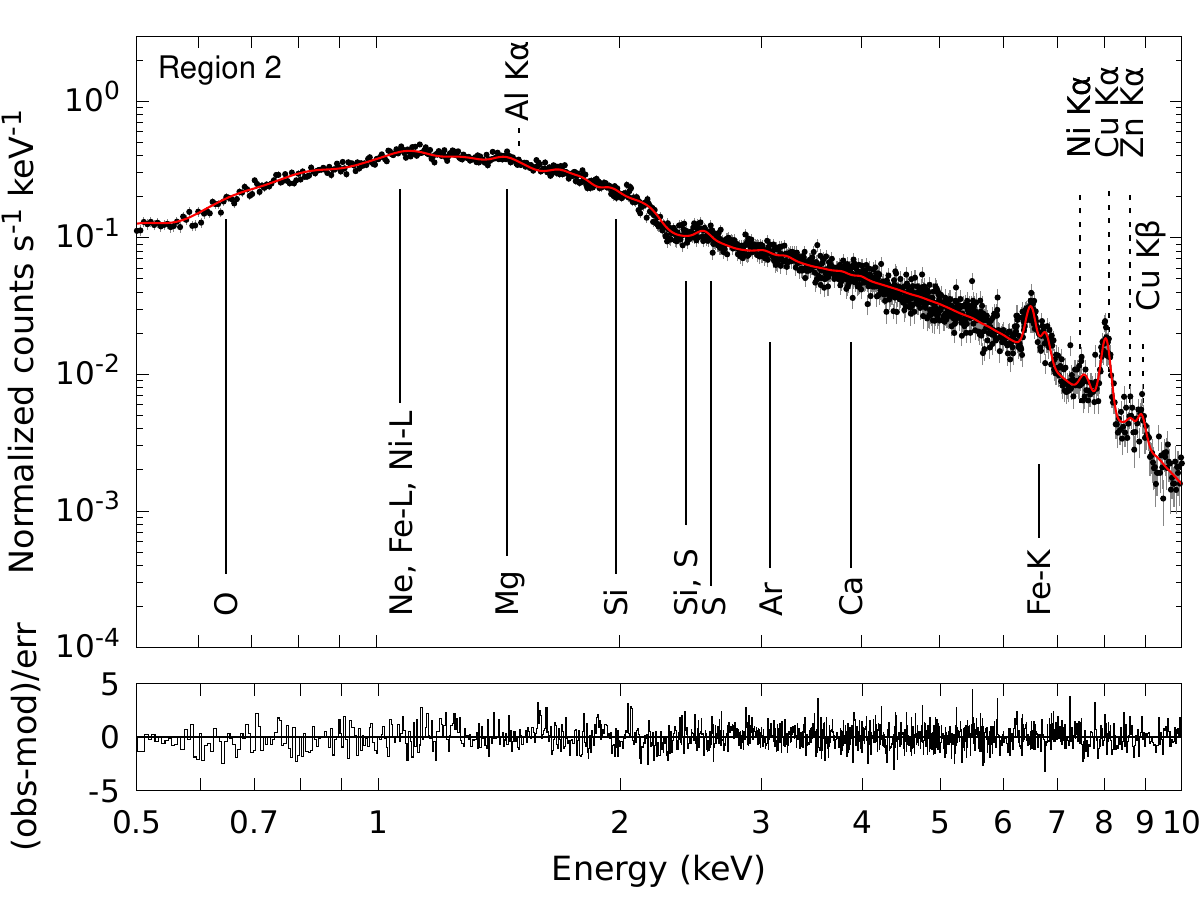} 
\includegraphics[width=0.33\textwidth]{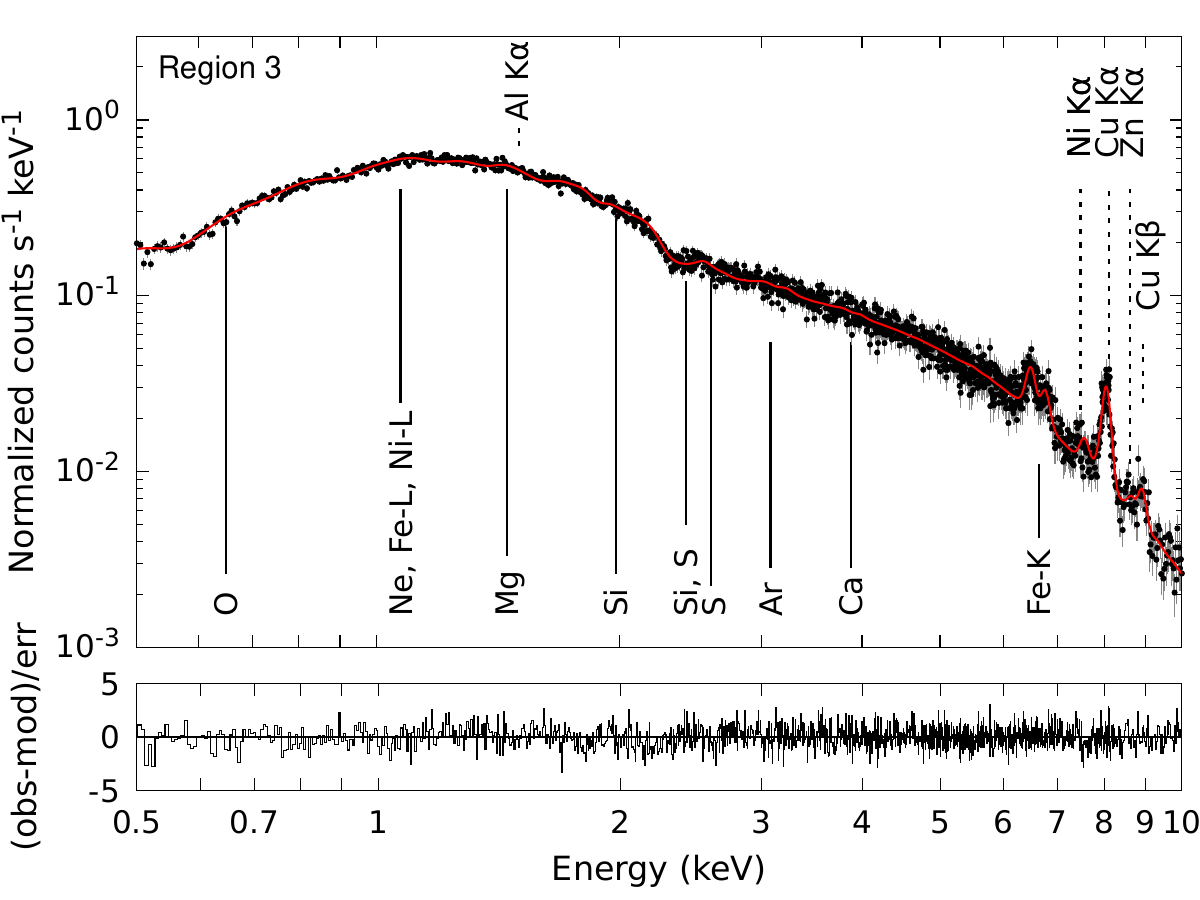}\\
\includegraphics[width=0.33\textwidth]{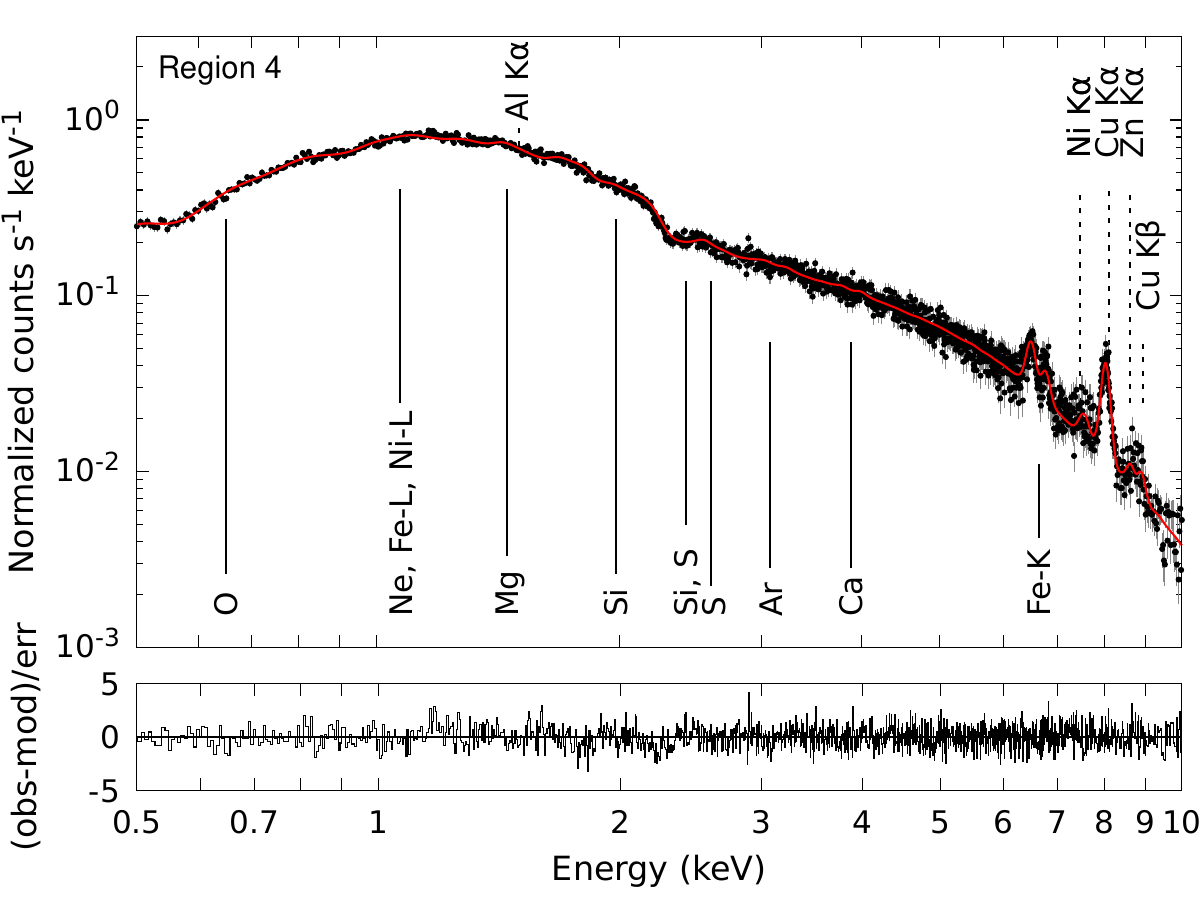} 
\includegraphics[width=0.33\textwidth]{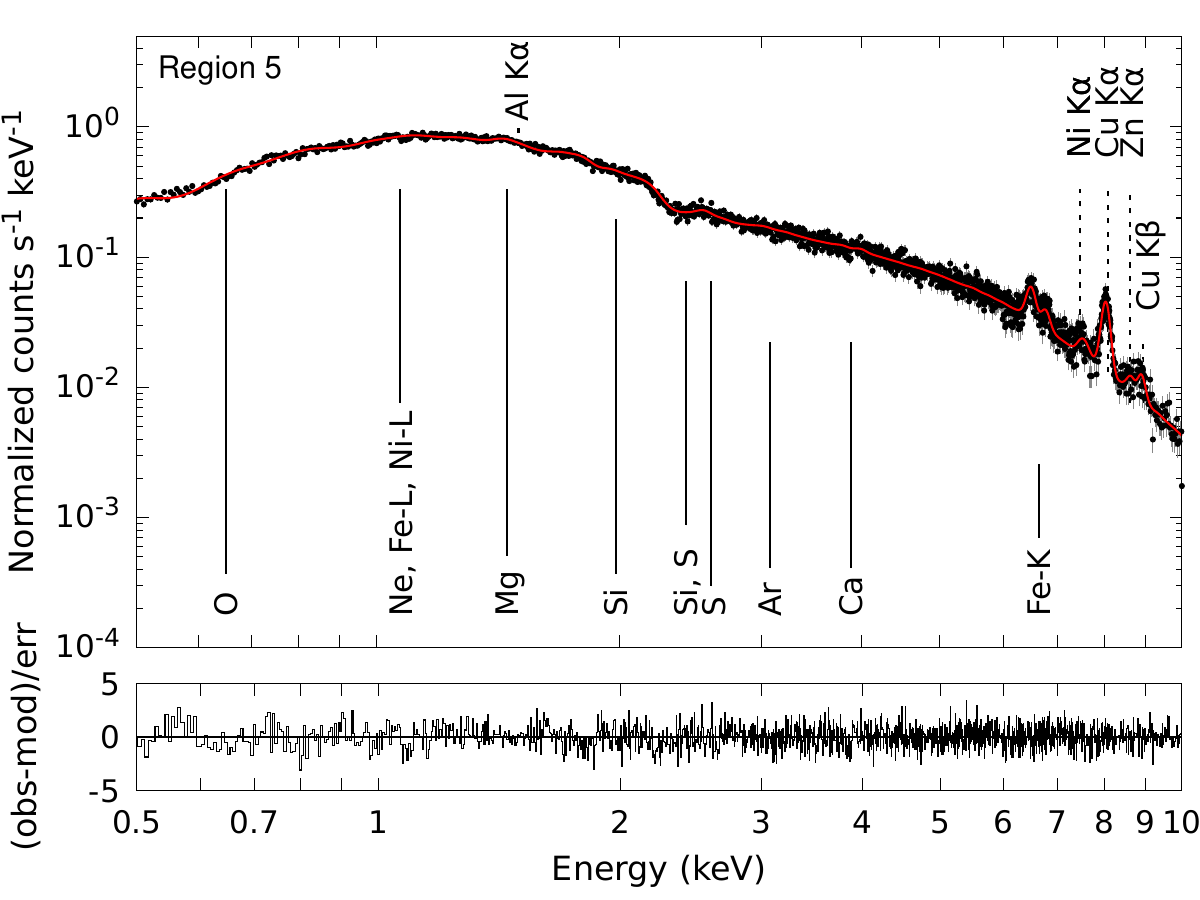} 
\includegraphics[width=0.33\textwidth]{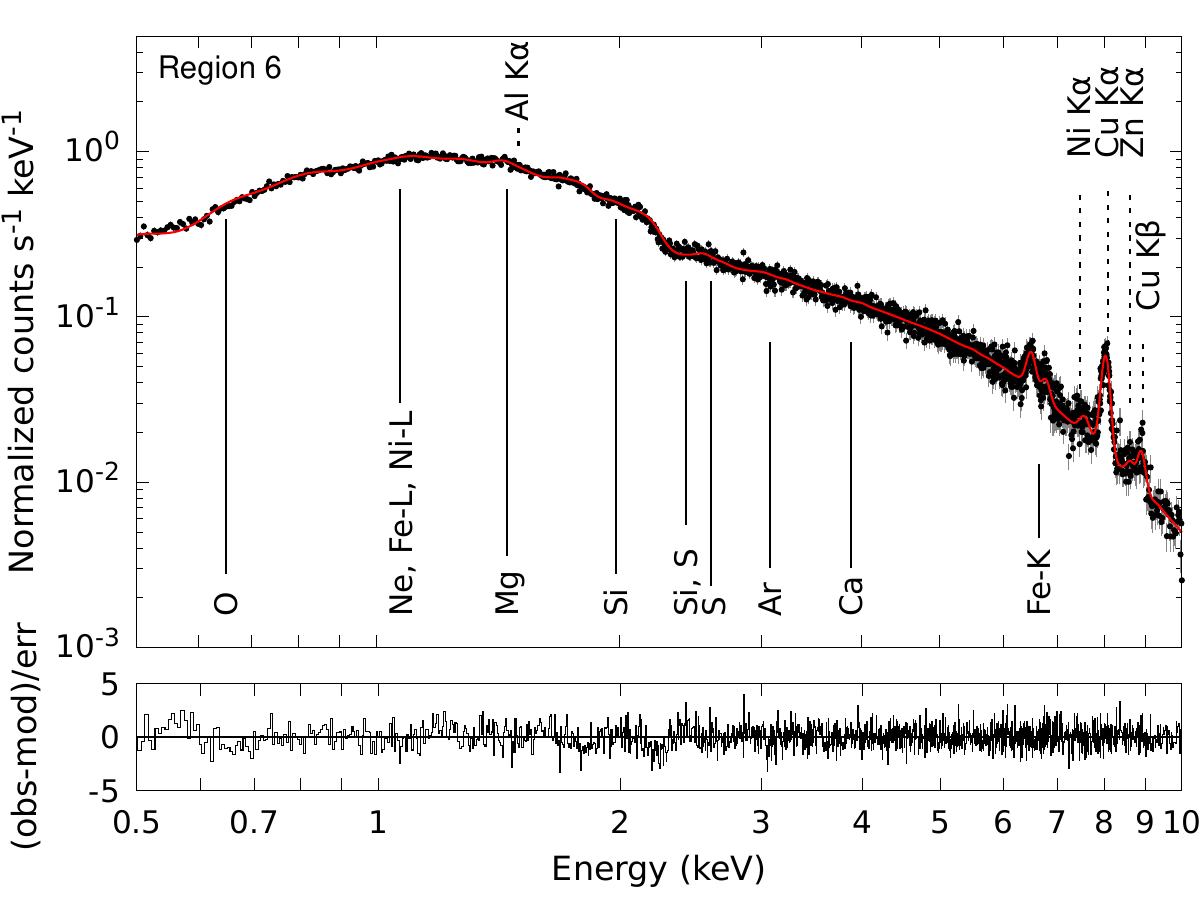}\\
\includegraphics[width=0.33\textwidth]{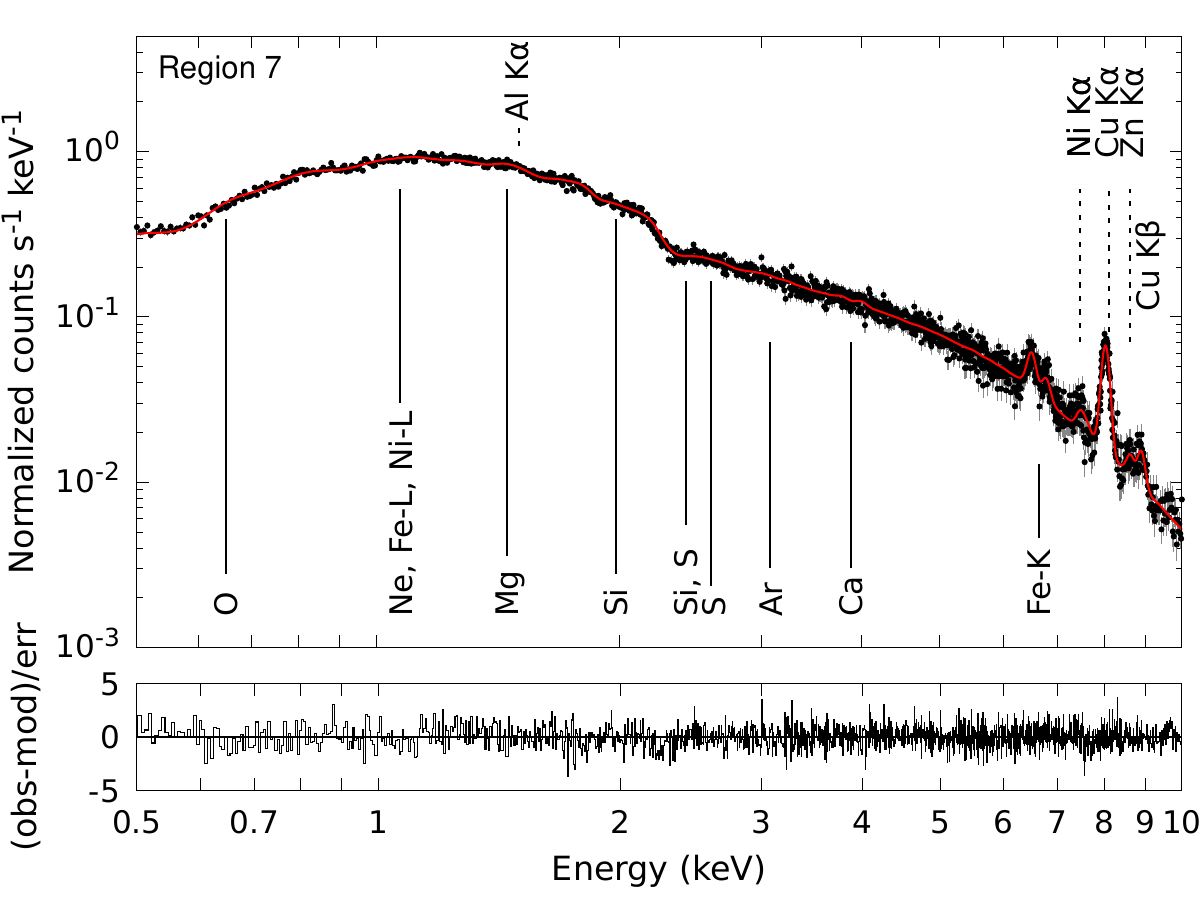} 
\includegraphics[width=0.33\textwidth]{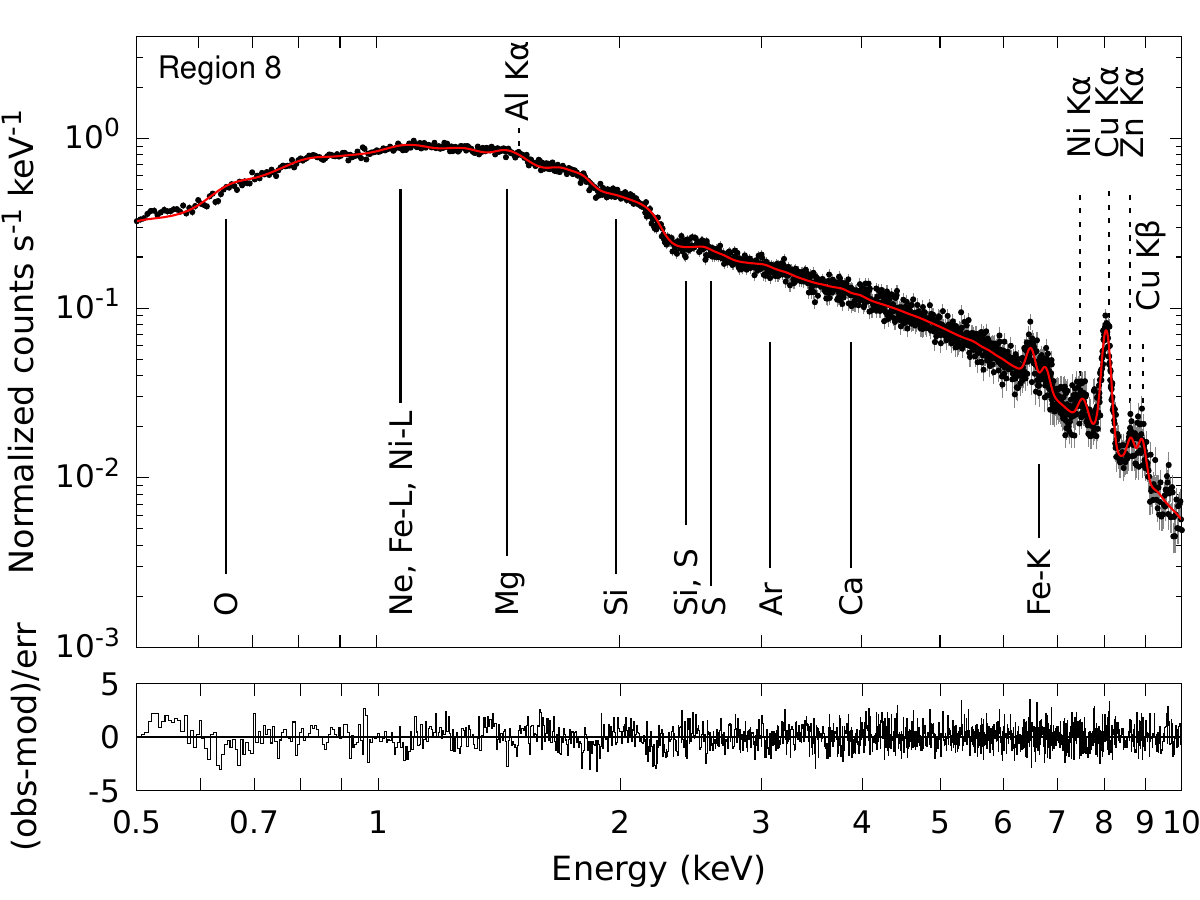} 
\includegraphics[width=0.33\textwidth]{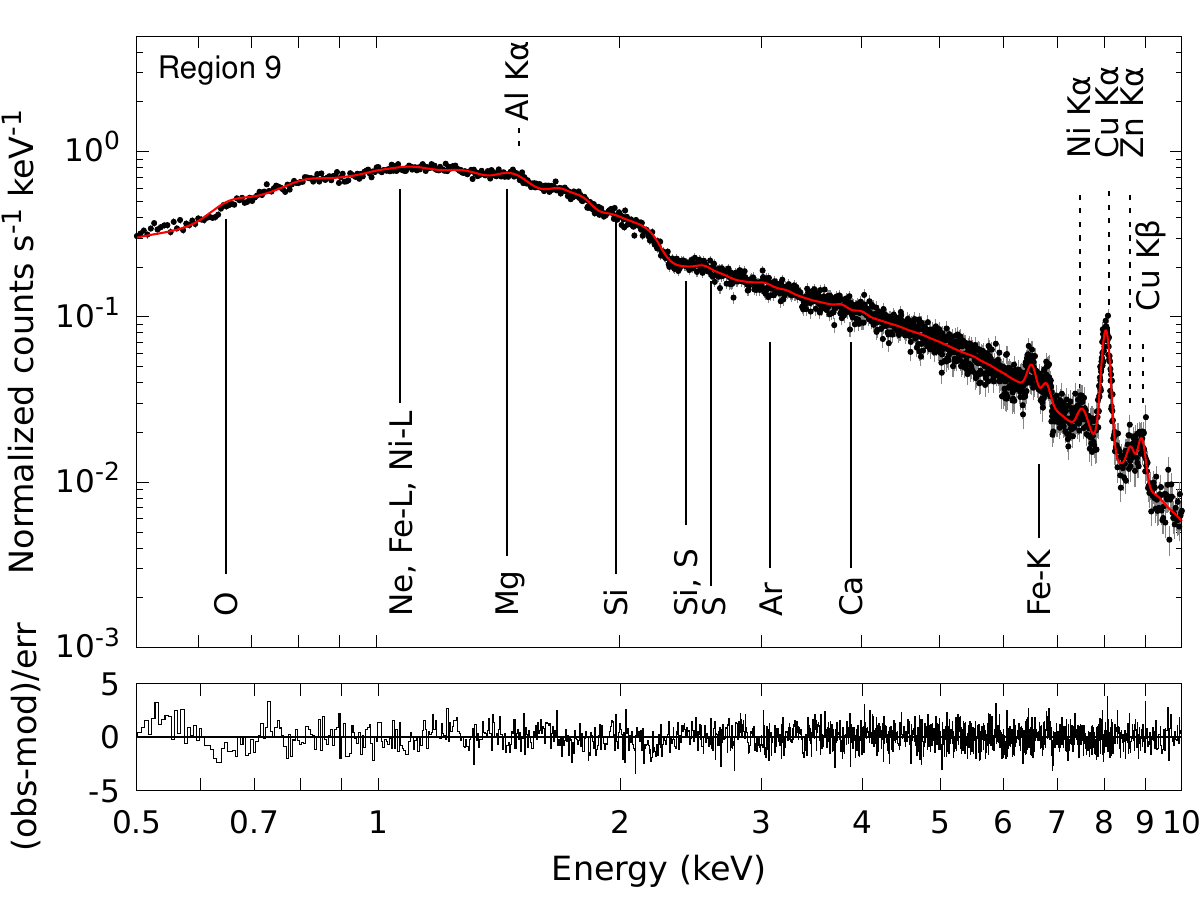}\\
\includegraphics[width=0.33\textwidth]{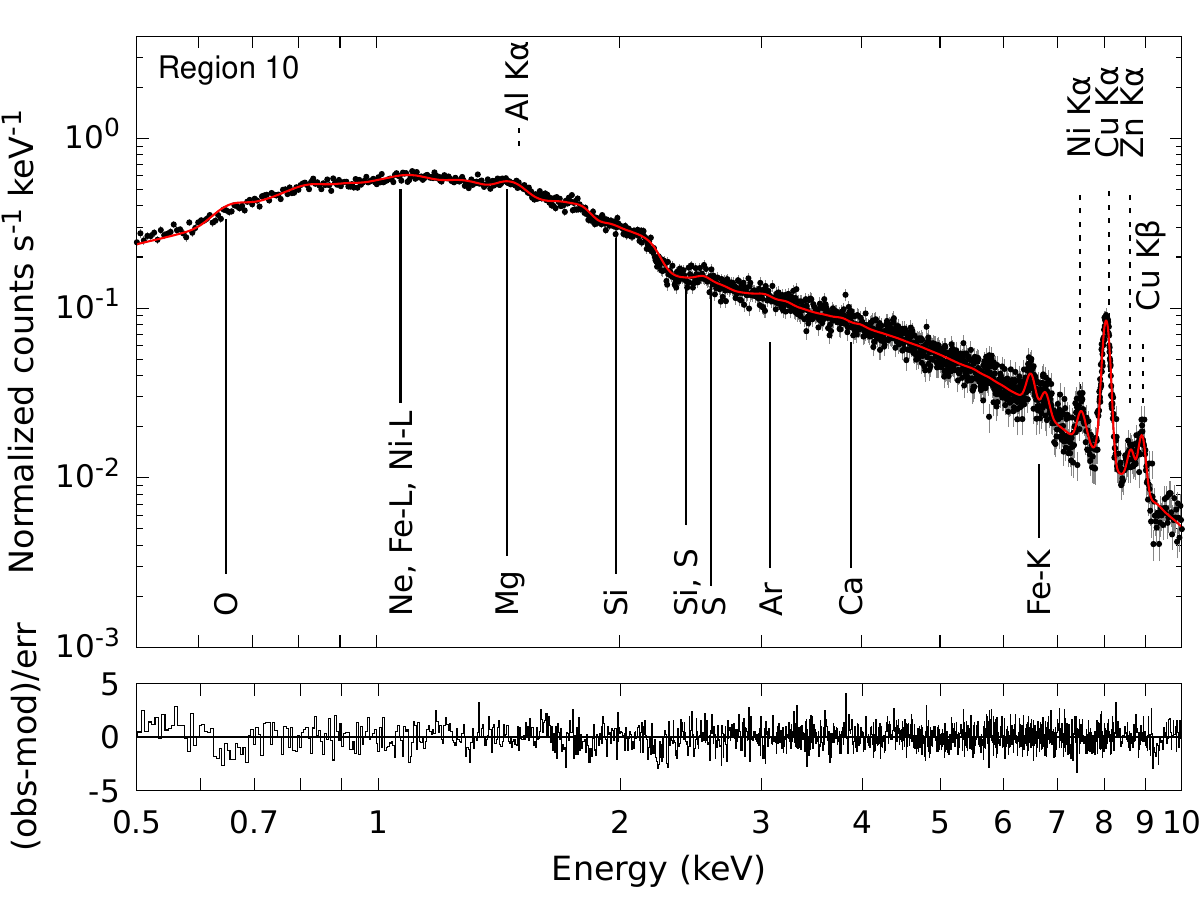} 
\includegraphics[width=0.33\textwidth]{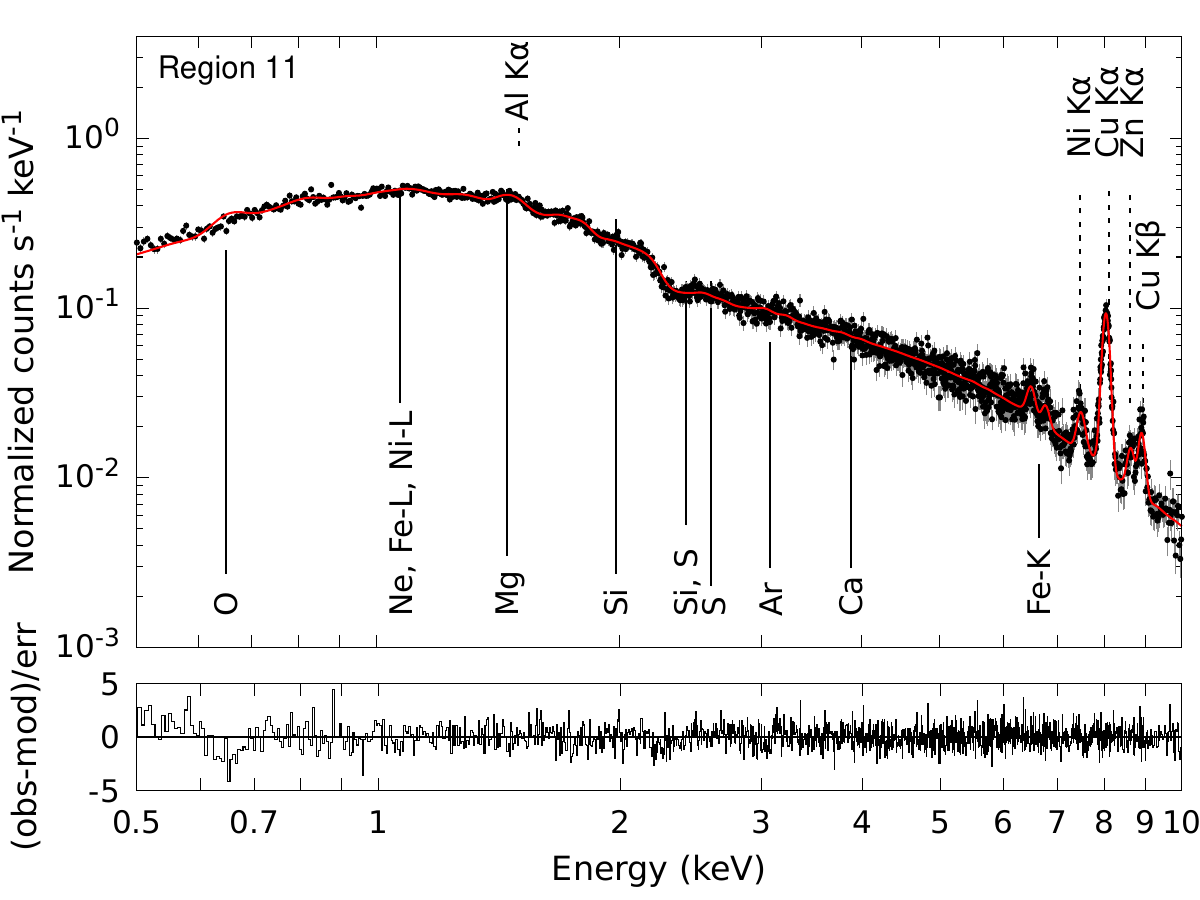} 
\includegraphics[width=0.33\textwidth]{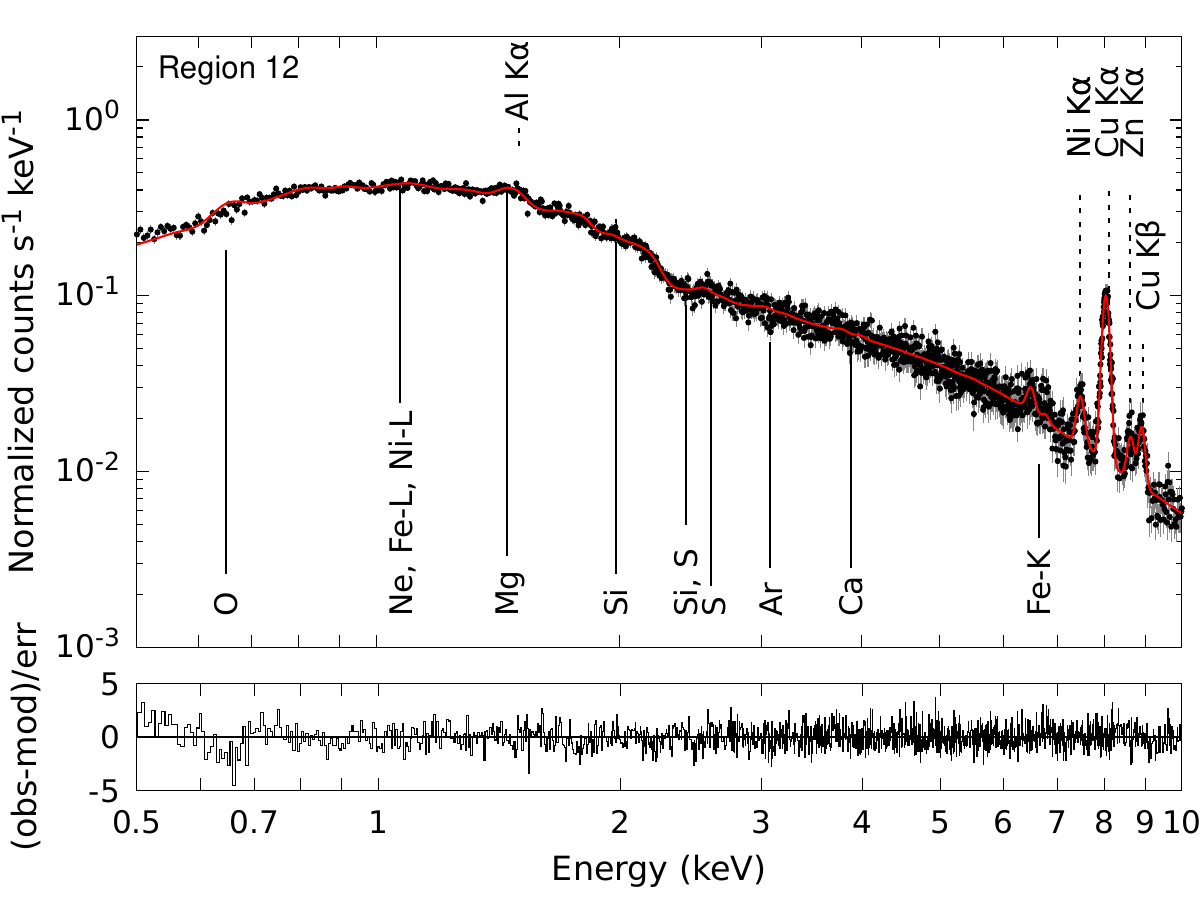}\\
\caption{
Best-fit spectra and model (red line) obtained for all regions analyzed within the Ophiuchus cluster. Vertical dashed lines indicate the line contribution from instrumental background while vertical solid lines indicate the ICM emission. Residuals are also included.  The spectra have been rebinned for illustrative purposes. The numbering is from the innermost to the outermost circle.
} \label{fig_spectra_example}
 
\end{figure*}

\section{Results and discussion}\label{sec_dis} 
Figure~\ref{fig_spectra_example} shows the best-fit spectra obtained in the 0.5-10~keV energy range using the model described above. In each panel, black data points correspond to the observation while the solid red line indicates the best-fit model for each case. The numbering is from the innermost to the outermost circle The base plots show the fit residuals. As a reference, we have included the instrumental Ni K$\alpha$, Cu K$\alpha$,$\beta$, Zn K$\alpha$, and Al K$\alpha$  background lines included in the model (vertical dashed lines) as well as the location of emission lines due to O, Ne, Mg, Si, S, Ar, Ca, Fe, and Ni from the ICM component. From the plots, it is clear that (1) the system is heavily absorbed and (2) the emission lines tend to be weak.

\subsection{Hydrogen column density}\label{sec_nh}  
The absorption column density in the line-of-sight to the Ophiuchus cluster is high due to its location near the Galactic plane, with a value of $N_{\rm H}=2\times 10^{21}$ cm$^{-2}$, and variations in the range of $N_{\rm H}=1.6-2.3\times 10^{21}$ cm$^{-2}$ in the fields centered within 1$^{\circ}$ of the Ophiuchus center \citep{kal05}. Moreover, 21~cm surveys of HI tend to underestimate the total absorption for systems located where molecular and dust absorption play an important role. This is particularly important in regions close to the galactic plane with abundant molecular clouds. According to previous analysis of the Ophiuchus cluster, treating $N_{\rm H}$ as a free parameter when modeling the ICM emission is crucial. This approach ensures that the molecular contribution and the variations across the observed sky region are adequately considered. Otherwise, unrealistically high temperatures would be obtained from the spectra fitting \citep{nev09,wer16,lov19}.  

\begin{figure}    
\centering
\includegraphics[width=0.45\textwidth]{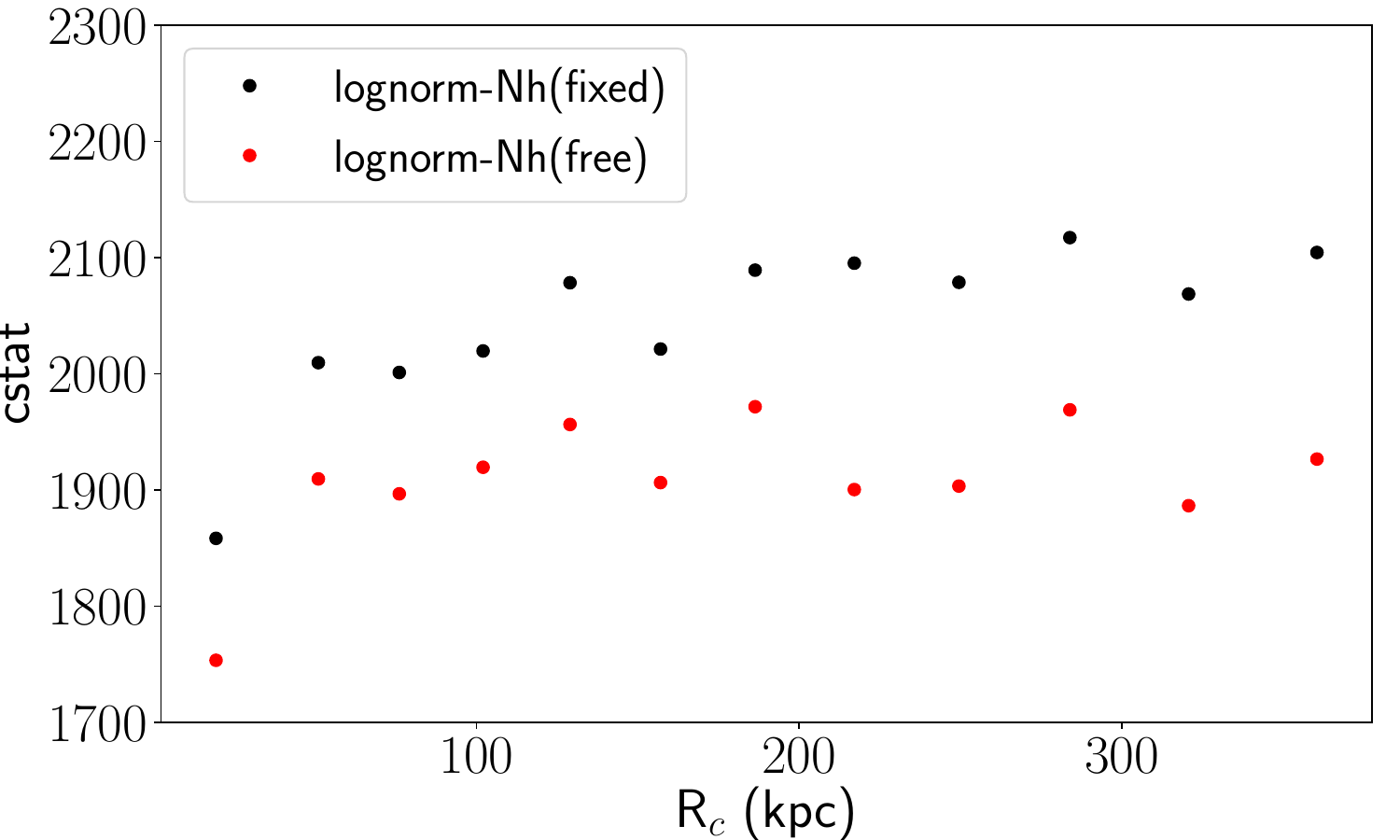}
\caption{
Best-fit cash statistic obtained with the {\tt lognorm} model with $N_{\rm H}$ as fixed and free parameters.
} \label{fig_cstat_comparison} 
\end{figure}

Figure~\ref{fig_kt_sigma} shows the hydrogen column densities obtained from the best-fit model (top panel). We found variations from $N_{\rm H}=3.68\pm 0.09\times 10^{21}$ cm$^{-2}$ to $N_{\rm H}=2.29\pm 0.07\times 10^{21}$ cm$^{-2}$, from the cluster core to the outermost region, leading to values almost two times higher than radio measurements, most likely due to the presence of molecular gas and dust. Furthermore, from dust extinction measurements, it has been estimated that $H_{2}$ contribution along this line-of-sight is substantial, up to $35\%$ of that of $H$ \citep{boh78,sch98,nev09}. Such molecular component contribution can impact larger than 10 percent in the measured metallicity \citep{sch15}.  Our best-fit results also agree with those reported by \citet{nev09}, who also found a decreasing profile with variations from $N_{\rm H}=3.50\pm 0.10\times 10^{21}$ cm$^{-2}$ to $N_{\rm H}=2.80\pm 0.10\times 10^{21}$ cm$^{-2}$. However, we are covering much larger distances. Finally, Figure~\ref{fig_cstat_comparison} shows the best-fit cash statistic obtained with the {\tt lognorm} model with $N_{\rm H}$ as fixed and free parameters. We found that the best-fit statistic corresponds to the model with $N_{\rm H}$ as a free parameter for all regions. Interestingly, recent works have discussed the Hidden Cooling Flows scenario where the cooled gas at the cool cores in galaxy clusters may collapse into very cold clouds, as well as substellar objects and low mass stars  \citep{fab22b,fab23a,fab23b}.

\subsection{Temperature profile}\label{sec_kt}  
\begin{figure}    
\centering
\includegraphics[width=0.45\textwidth]{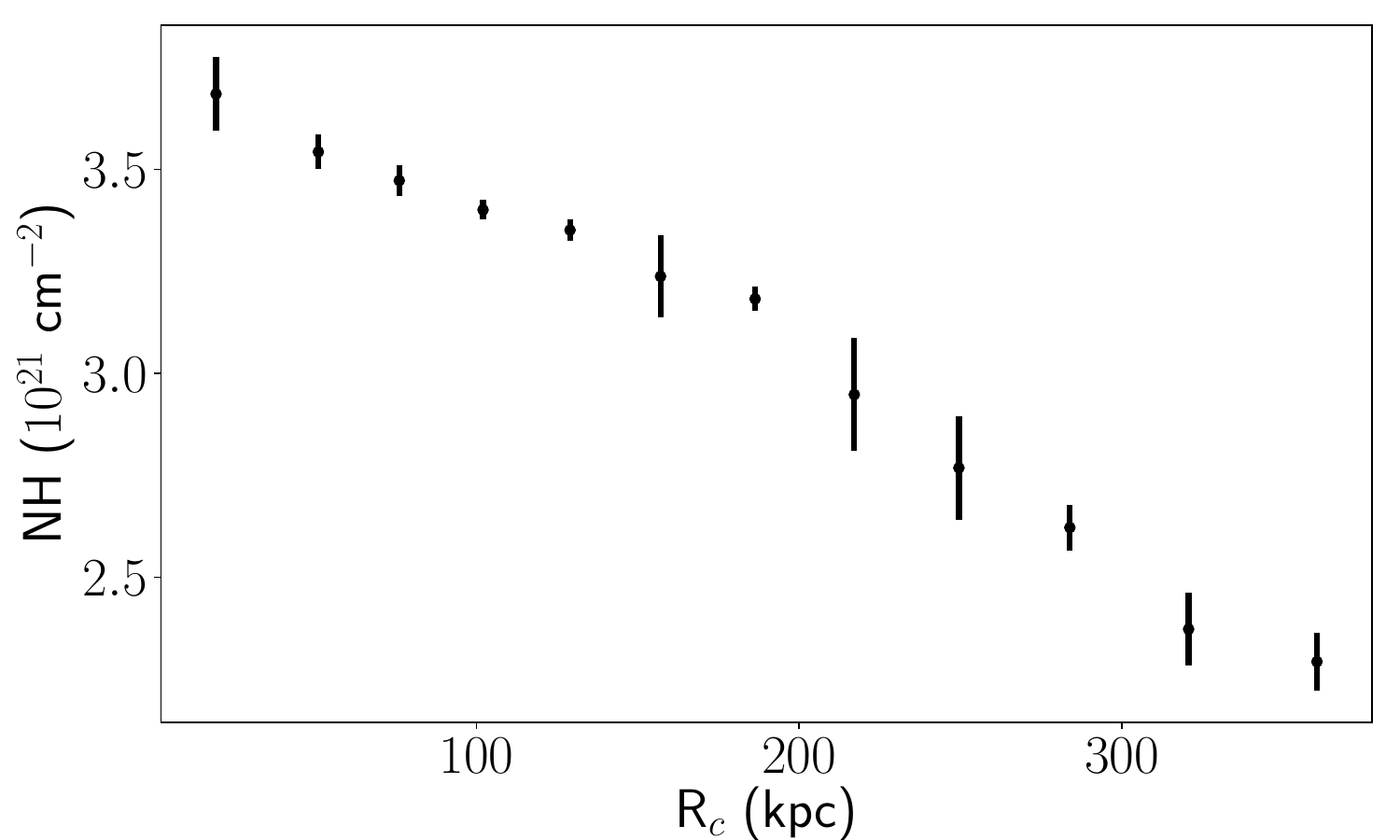}\\
\includegraphics[width=0.45\textwidth]{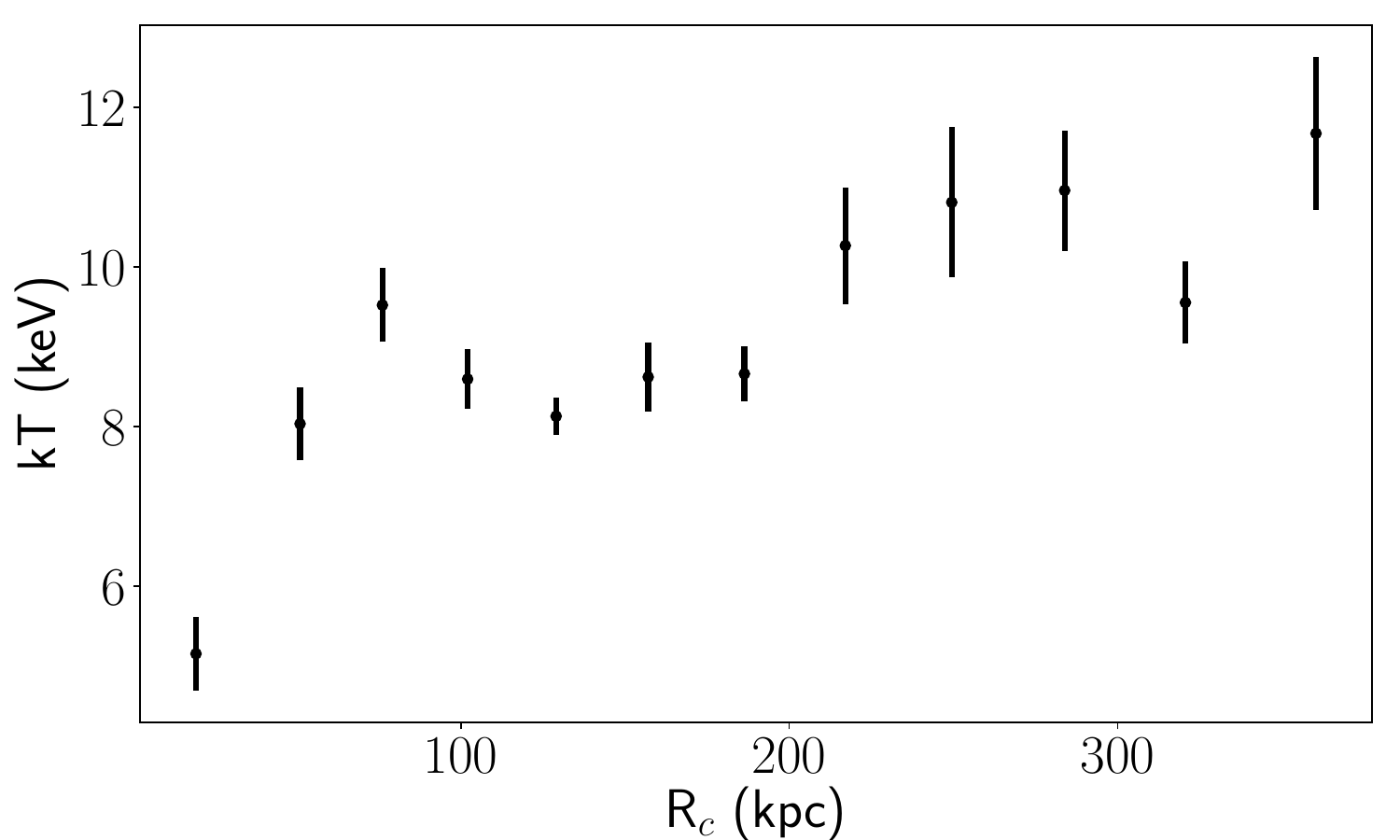}\\
\includegraphics[width=0.45\textwidth]{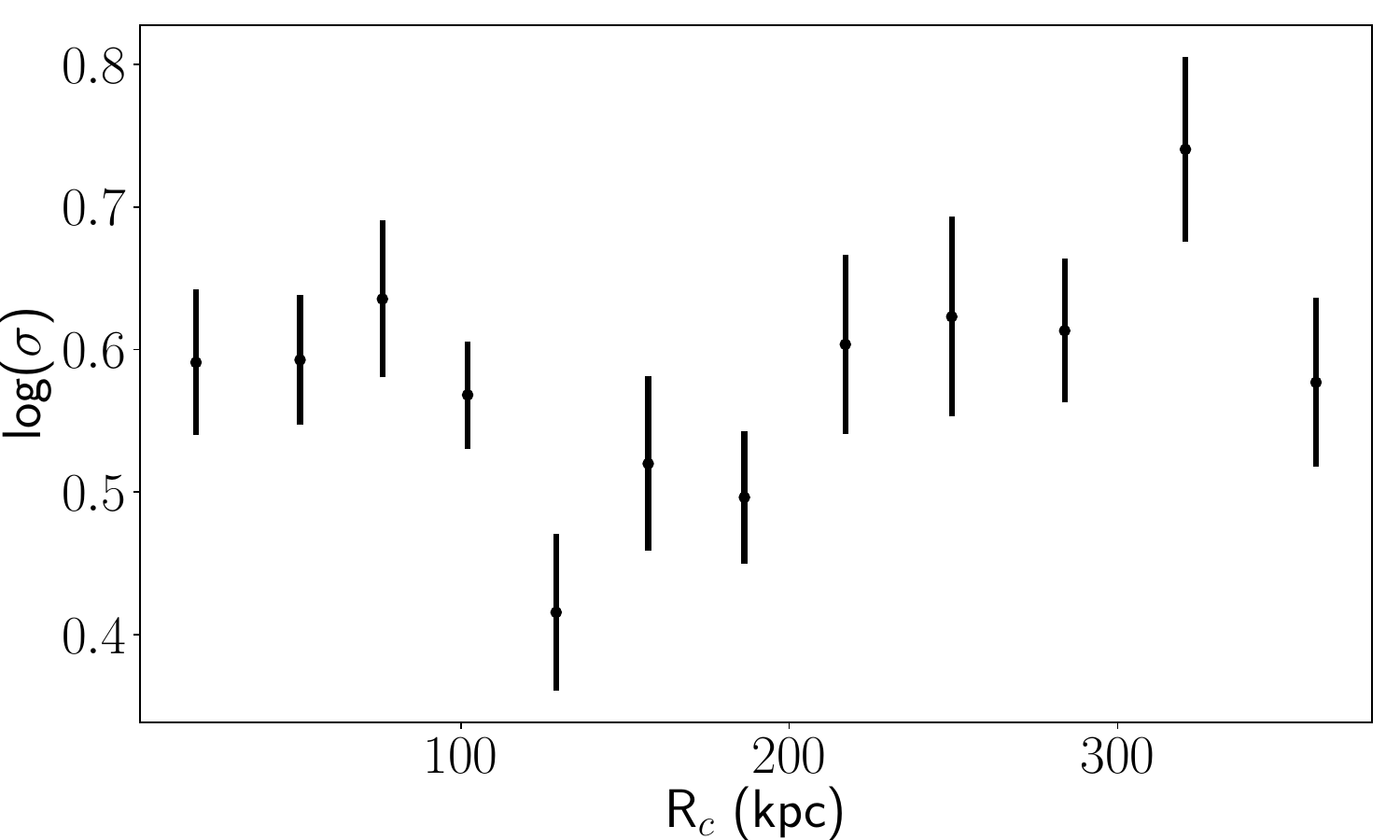}
\caption{
Best-fit values obtained for the hydrogen column density (top panel), temperature (top panel) and $\log(\sigma)$ (bottom panel) as function of distance to the Ophiuchus cluster center. Note that the points are located in the middle of the extraction rings (i.e. region 1 covers from 0 to $32$~kpc).
} \label{fig_kt_sigma} 
\end{figure}

Figure~\ref{fig_kt_sigma} shows the best-fit values obtained for the temperature ($kT$, middle panel) and $\log(\sigma)$ (bottom panel). The peak temperature of the distribution increase from $5.5\pm 0.46$~keV to $11.67\pm 0.95$~keV (for regions 1 and 12, respectively), while the $\log(\sigma)$ parameter varies from $0.41\pm 0.05$ to $0.74\pm 0.06$ (for regions 5 and 11, respectively). These temperatures are in the same range found by \citet{nev09,gat23a}, although they both used a single-temperature model. The temperature profile heavily increases for distances $< 100$~kpc and then continues increasing at a slower rate. Previous works have shown a relatively isothermal temperature distribution \citep{mil10}. The $\log(\sigma)$ profile indicates that the multi-temperature model is required for all regions.

\subsection{Velocity profile}\label{sec_vel}  
\begin{figure}    
\centering  
\includegraphics[width=0.45\textwidth]{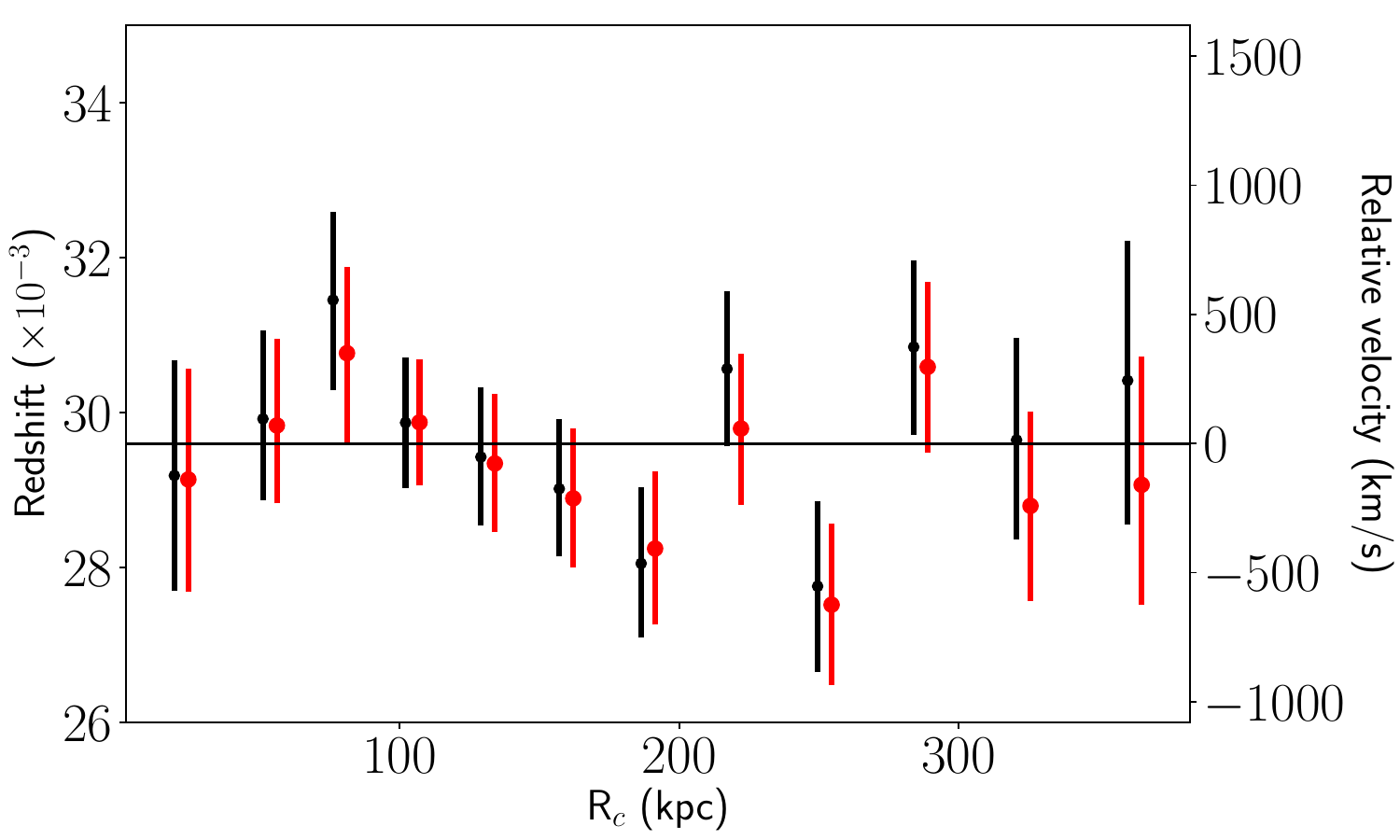}  
\caption{ 
Velocities obtained for each region (black points) and those obtained by \citet[][red points]{gat23a}. The Ophiuchus cluster redshift is indicated with an horizontal line. Note that the points are located in the middle of the extraction rings (i.e. region 1 covers from 0 to $32$~kpc).
}\label{fig_vel} 
\end{figure}

The best-fit velocities obtained for each region are shown in Figure~\ref{fig_vel}. Black points correspond to the values obtained in the present analysis while red points correspond to the values obtained by \citet{gat23a}. The results agree even though we include the soft band $<4$~keV and $N_{\rm H}$ as a free parameter in the modeling. This indicates the robustness of the velocity measurement method. In particular, we notice the little impact the multitemperature modeling and elemental abundances as free parameters have in the iron complex redshift measurements obtained with the EPIC-pn camera. We found that most of the velocities show no deviation from the cluster velocity. The largest redshift/blueshift with respect to Ophiuchus is $477\pm 330$ and $-586\pm 450$ for rings 3 and 9, respectively. This corresponds to a departure from the system velocity at about $2\sigma$ confidence level. The mean and error-weighted standard deviation for the distribution is $\mu=15$~km/s and $\sigma=314$~km/s, respectively.

\subsection{Abundance profiles}\label{sec_abun} 
\begin{figure}    
\centering  
\includegraphics[width=0.415\textwidth]{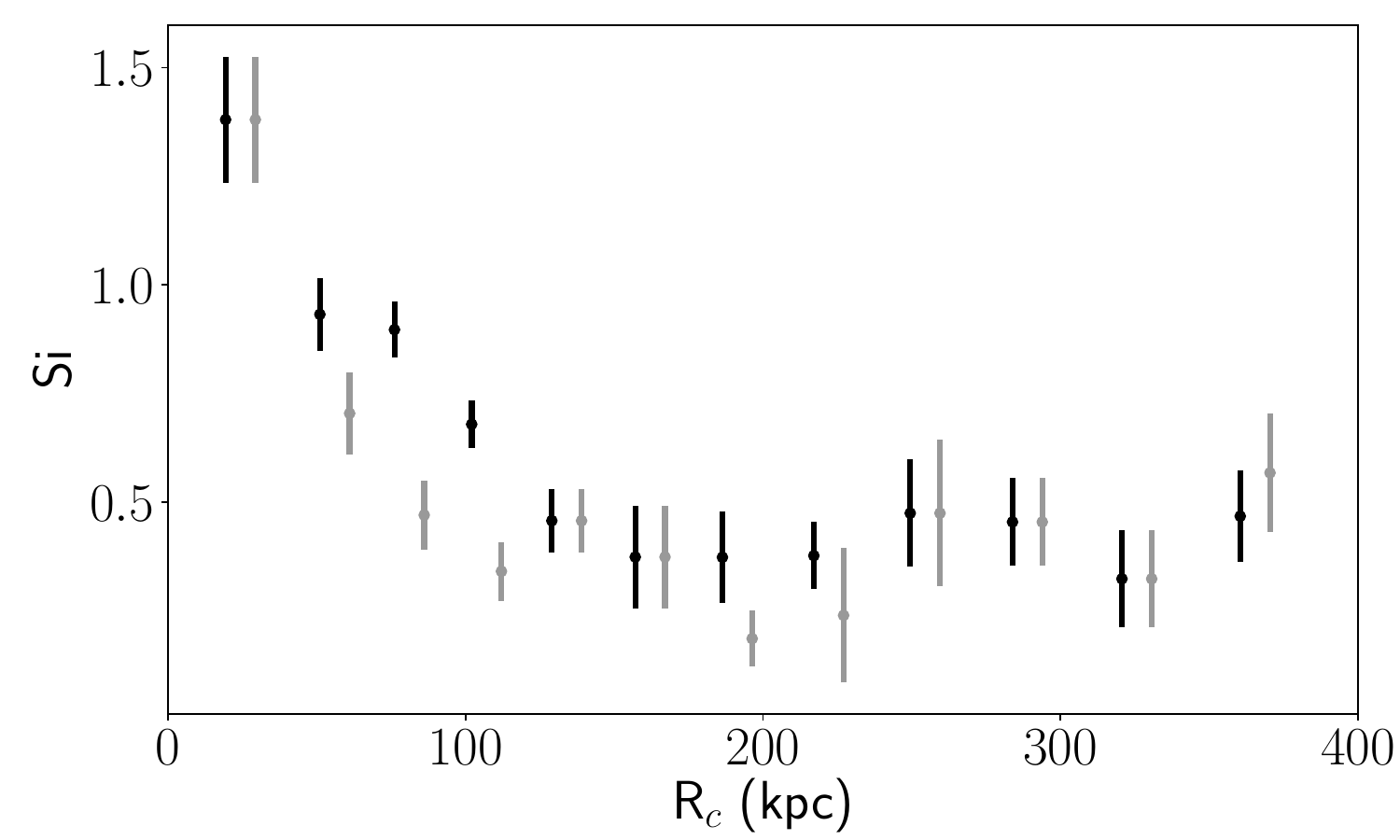}
\includegraphics[width=0.415\textwidth]{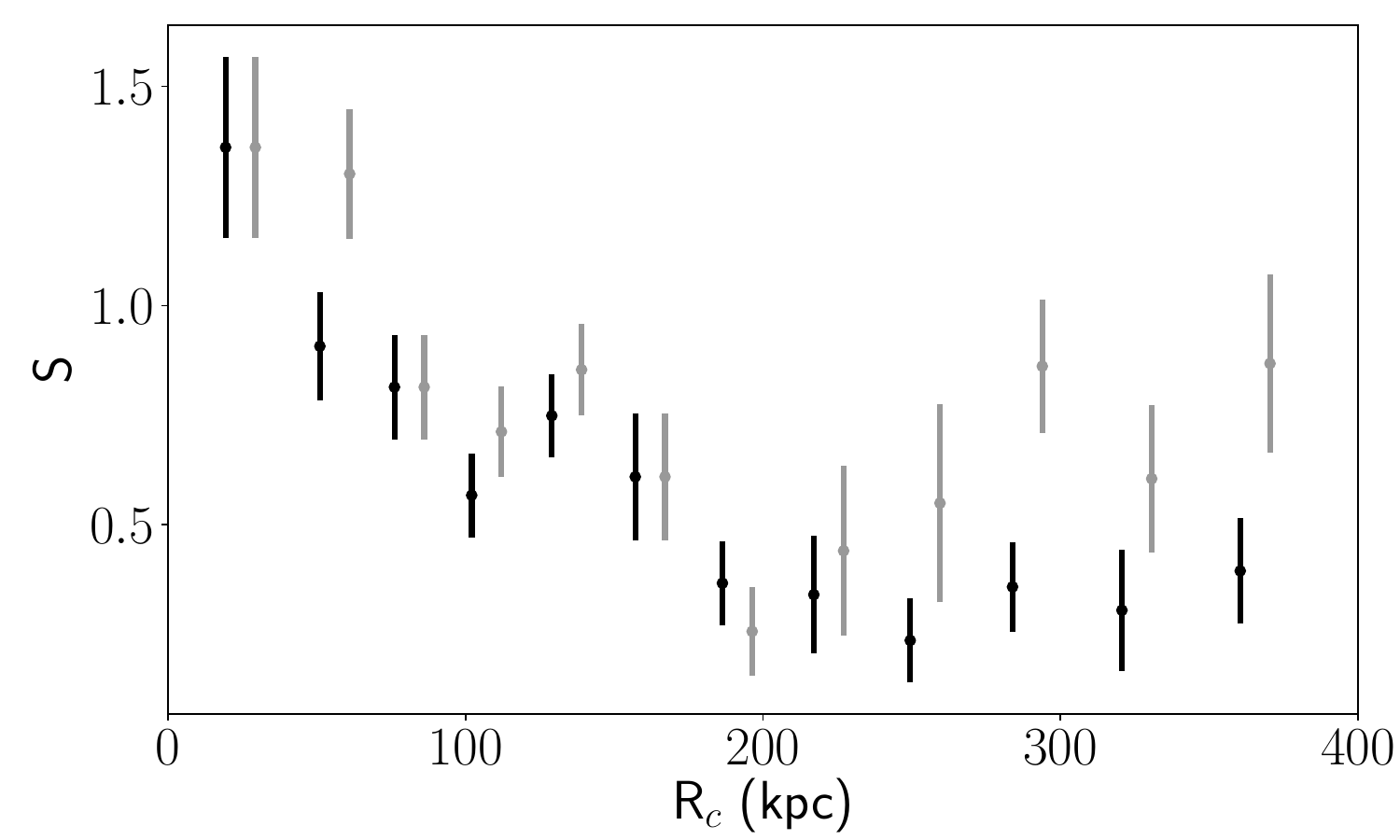}\\
\includegraphics[width=0.415\textwidth]{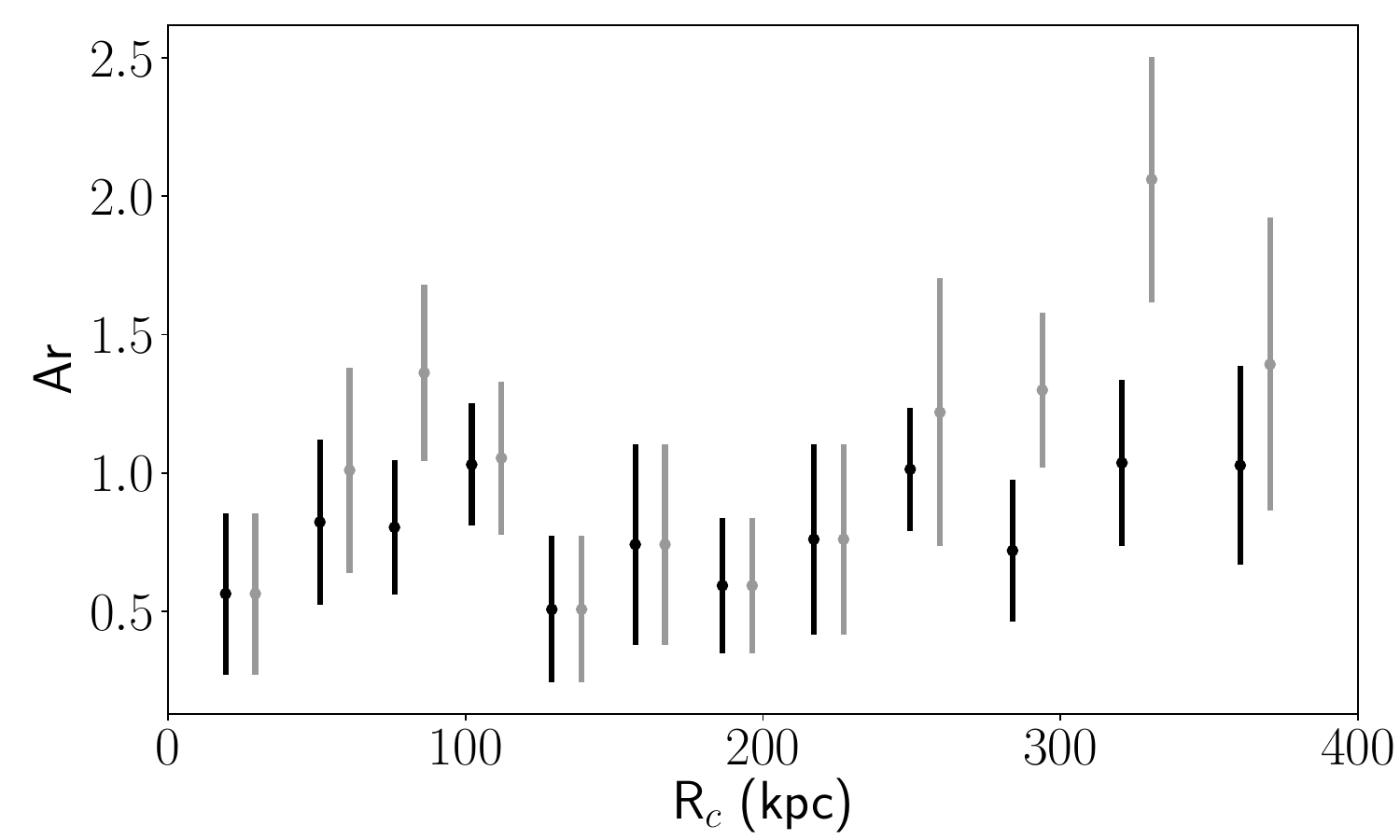}\\
\includegraphics[width=0.415\textwidth]{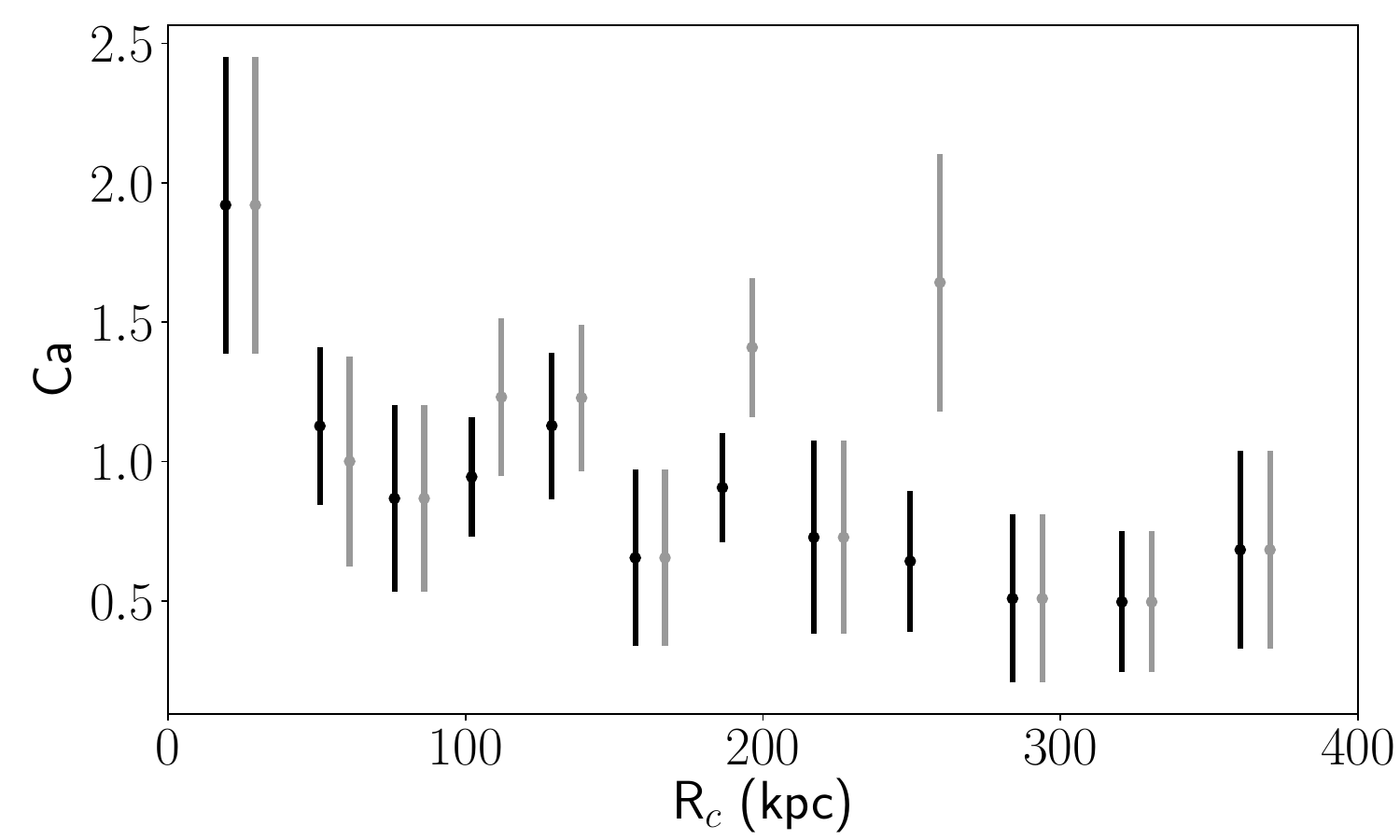} \\
\includegraphics[width=0.415\textwidth]{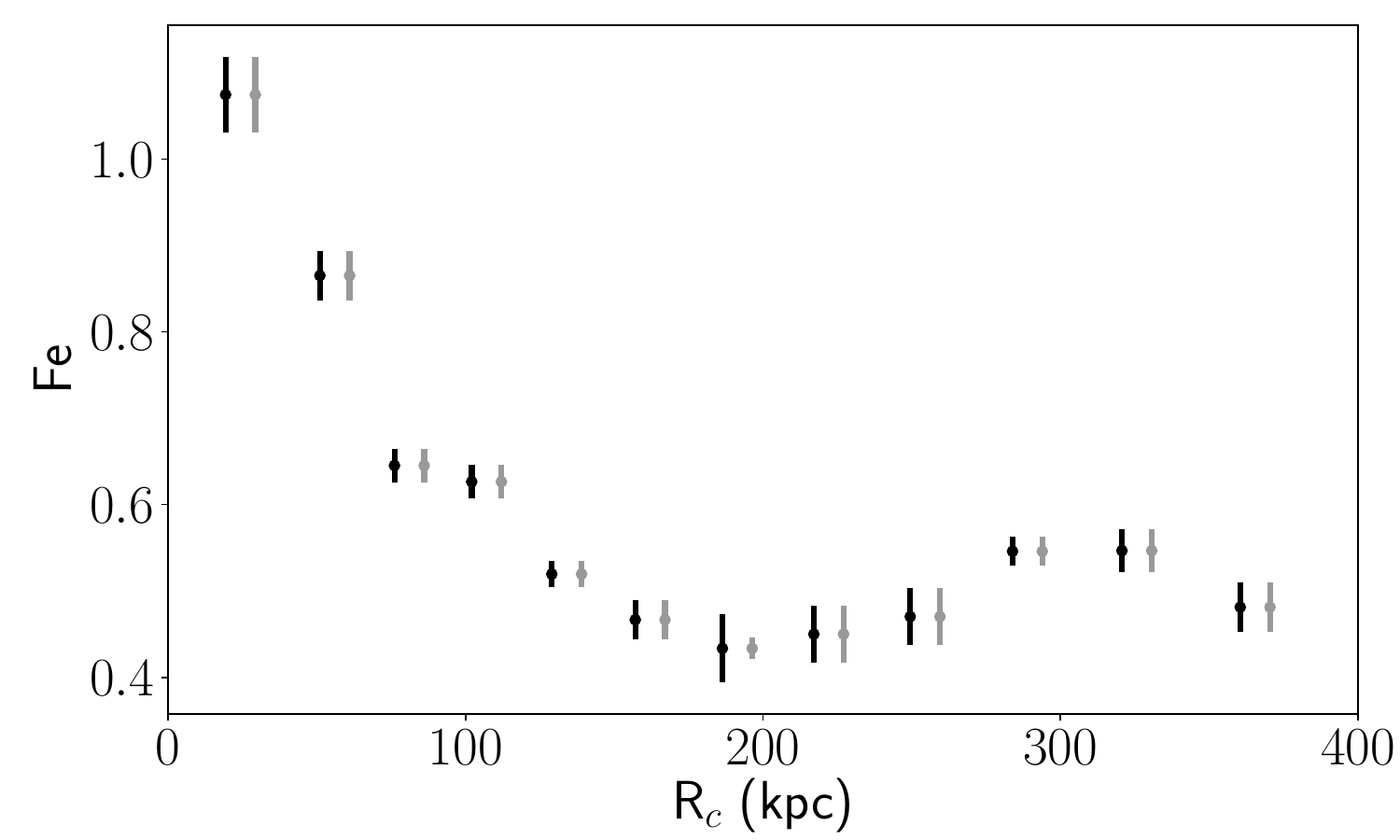}\\  
\caption{ Abundance profiles obtained from the best-fit results for the Ophiuchus cluster. Note that the points are located in the middle of the extraction rings (i.e. region 1 covers distances from 0 to $32$~kpc). Black points correspond to abundances obtained from local fits while gray points correspond to those obtained from the global fits (see Section~\ref{sec_fits}). 
} \label{fig_abund_all} 
\end{figure}

The best-fit elemental abundances obtained with respect to the solar values are shown in Figure~\ref{fig_abund_all}. For O, Ne, Mg, and Ni we have obtained only upper limits for most of the regions. This is expected because the Ophiuchus cluster is heavily absorbed, therefore it is difficult to observe the emission lines corresponding to these elements. (see Figure~\ref{fig_spectra_example}). We have obtained a clear decreasing profile for Fe as a function of the distance to the cluster center. While Fe abundances are well constrained due to the presence of the Fe K$\alpha$ complex, the uncertainties for other elements are large. Furthermore, we have not found changes in the abundance profiles with statistical significance $>2\sigma$, except for a notable decrease of Si and S abundances for distances $<50$~kpc before becoming somewhat flatter. While correlations with velocities are not found, we have found that the cooler gas is more iron-rich when comparing the abundance profiles with the temperature profile. Given the high temperatures found within this system, it is possible that the low strength of the emission lines is due to the high ionization state of the ICM. Figure~\ref{fig_abund_all} also include the abundances obtained from the global fits (gray points, see Section~\ref{sec_fits}). The local fits give better constraints in the abundances and reduce the measurements scattering, especially for larger distances. In the following analysis, we consider the abundances obtained from the local fits.

A central increase of metal abundance was identified in previous studies \citep{fuj08,nev09}. However, we note that the spatial scale of the innermost region analyzed is much larger than the physical radius for which a drop in Fe abundance has been identified previously \citep[i.e., $<5$~kpc scale. See ][]{liu19}. Also, metal-rich regions at large distances (e.g. increasing of iron for distances $>200$~kpc, see Figure~\ref{fig_abund_all}) could be structures that have been stripped from the cool core during its motion \citep{mil10,wer16}. Future comparisons with detailed hydrodynamical simulations including cluster merging events are important to better understand the dynamics and kinematics within such a system.

\citet{liu19} obtained abundance profiles for Fe, Ar, and Ne in their analysis of {\it Chandra}  observations, covering up to 45~kpc distance from the cluster core. Our results for Fe and Ar are in good agreement with theirs, although the spatial scale in our study is larger. Similarly, the Fe profile agrees with \citet{wer16}. While there are few analyses of individual abundances for such a hot cluster, there have been numerous studies on Fe abundances in other massive clusters. \citet{mer17} measured radial metal abundance profiles in the ICM using {\it XMM-Newton} EPIC-pn observations for multiple galaxy clusters. For the Fe abundance, they found a peak in the core and an overall decrease of the Fe abundance with the radius out to $\sim 0.9$ $r_{500}$ in clusters, a distribution predicted from hydrodynamical simulations \citep{pla14}. A significant drop in the Fe abundance was also identified in $39\%$ of their sample. Such drop in metallicity, within distances of $<10$~kpc from the cluster center, has been identified in other clusters \citep{san02,pan13,san16}, although the origin of such a drop is a subject of extensive debate \citep[see for example][]{fuk22}. In that sense, we have analyzed regions larger than the drop scale for the innermost region of the cluster. More recent work on abundance profiles in galaxy clusters was done by \citet{sim21} using the X-COP sample \citep{eck17}. They also found abundance profiles flattening out at large radii with an average level of $0.39$ solar abundance for hot clusters \citep[i.e., scaled to solar abundance from ][]{lod09}. Our measured Fe profile shows higher metallicity at large regions ($\sim$0.45 solar abundance, see Figure~\ref{fig_abund_all}). In this aspect, the mixing of metal-rich gas in the outskirts may be due to complex merger activities \citep{sun02,urd19,wal22}.

\subsection{ICM chemical enrichment from SN}\label{sec_snr} 
\begin{figure}    
\centering   
\includegraphics[width=0.45\textwidth]{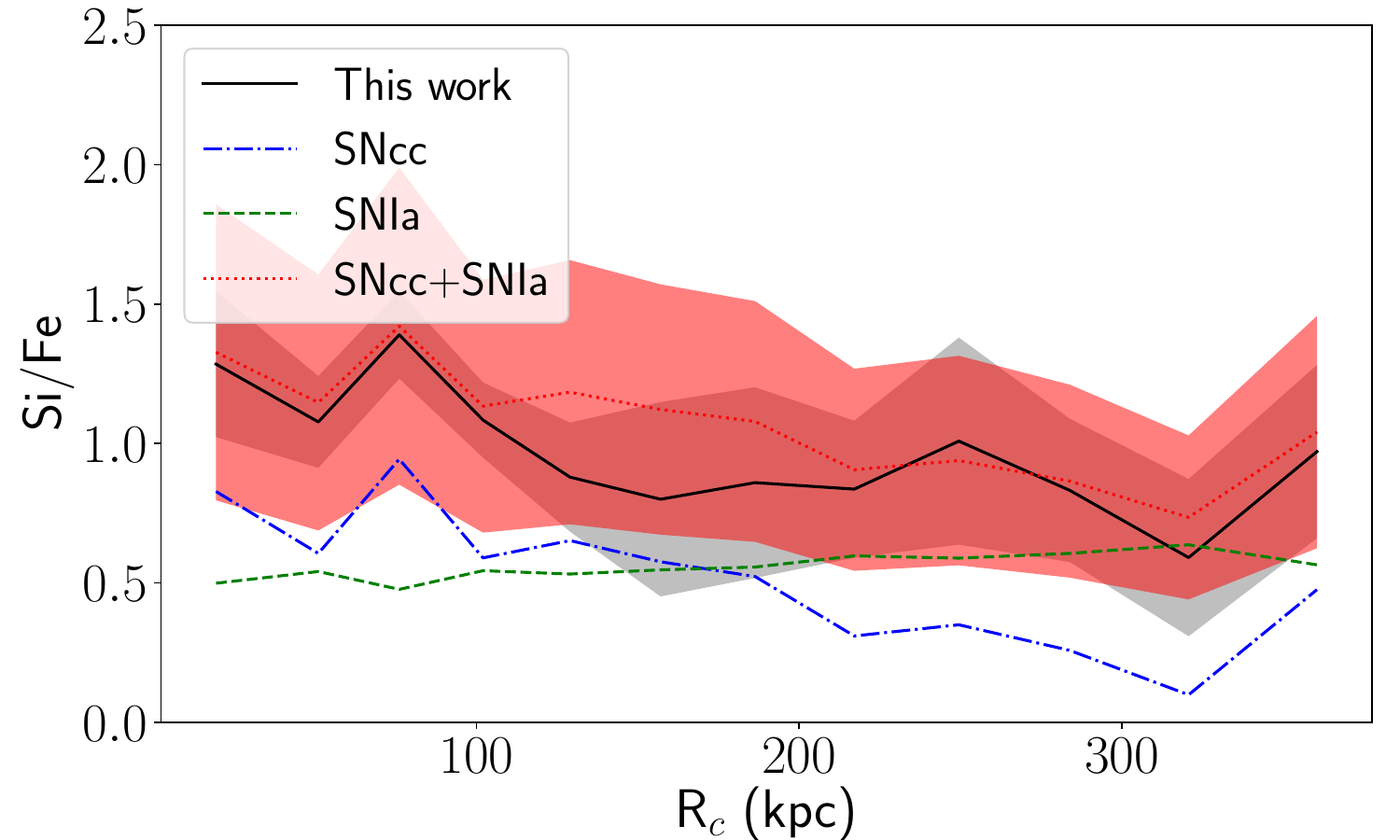}  \\
\includegraphics[width=0.45\textwidth]{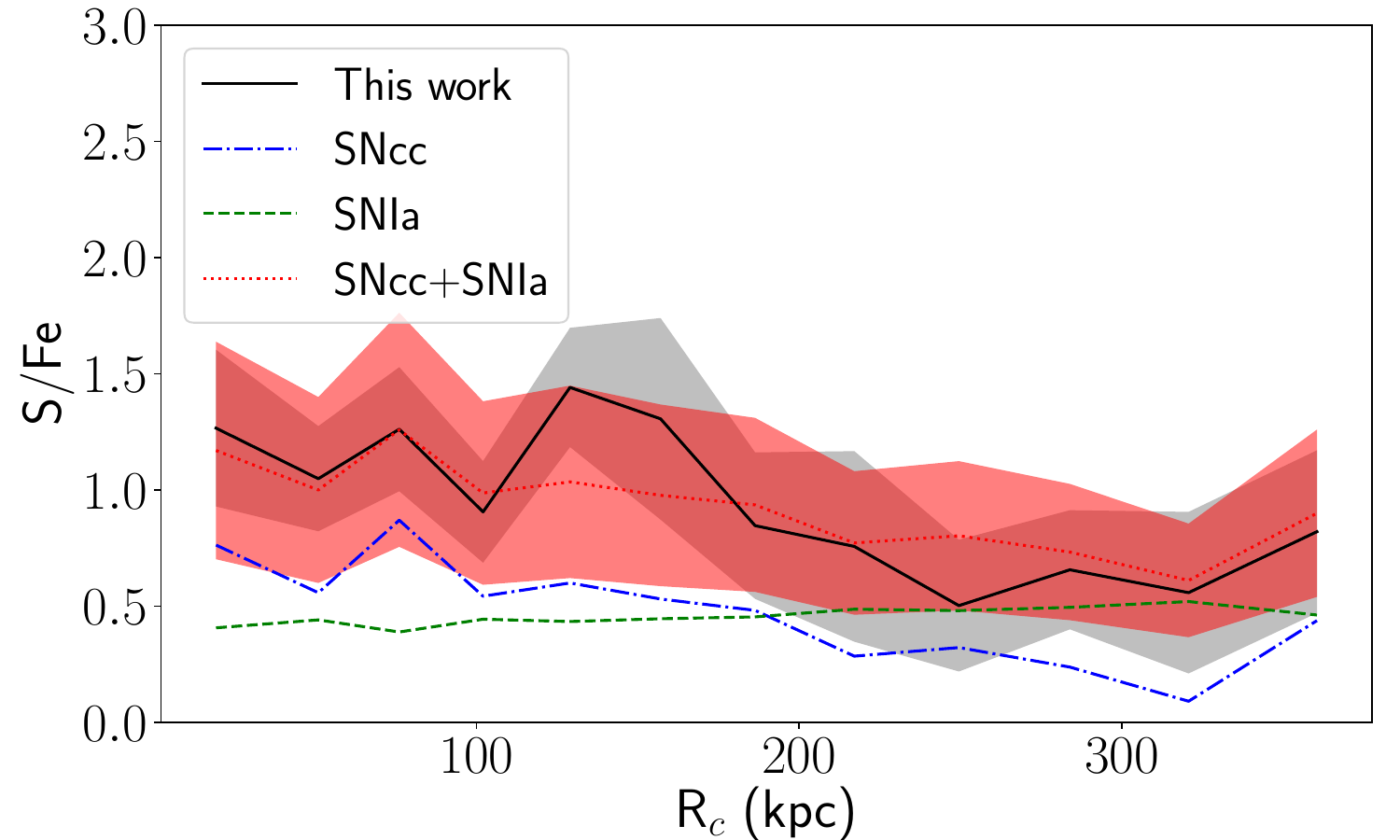}  \\
\includegraphics[width=0.45\textwidth]{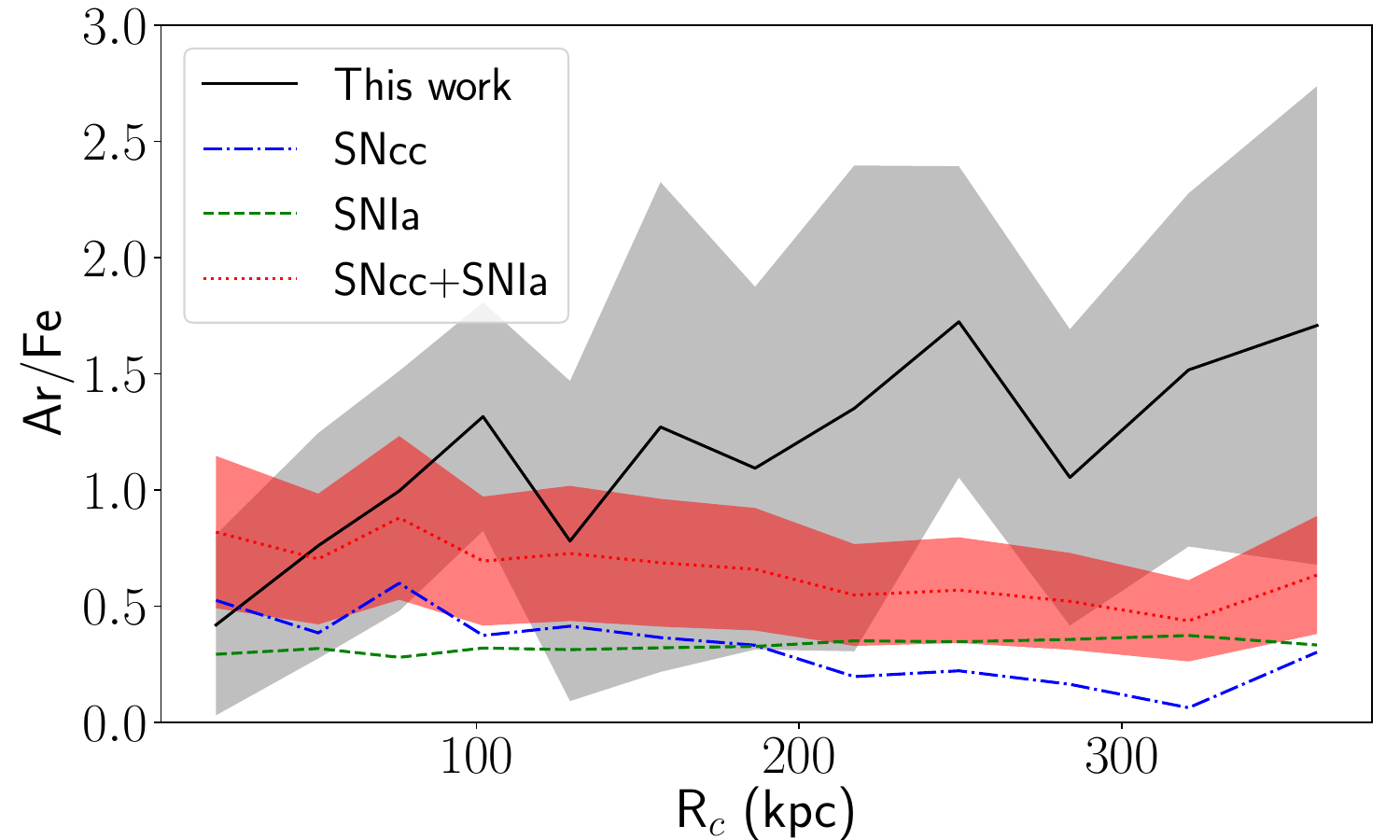} \\
\includegraphics[width=0.45\textwidth]{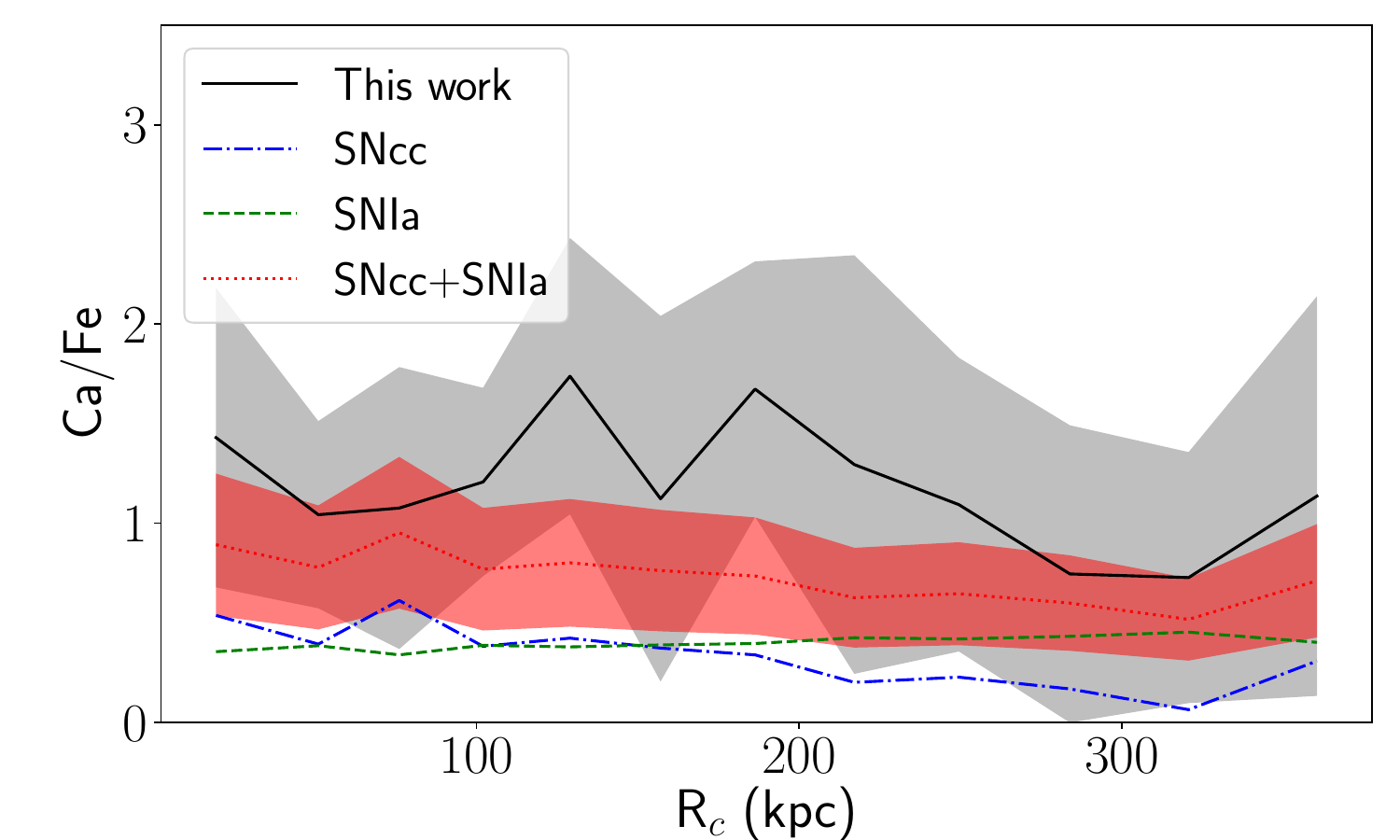}   
\caption{
Abundance ratio profiles, relative to Fe, obtained from the best-fit results for the Ophiuchus cluster. The gray shaded areas indicate the mean values and the 1$\sigma$ errors. The SNcc (blue line), SNIa (green line) contribution to the total SN ratio (red shaded area) from the best fit model are included (see Section~\ref{sec_snr}). 
 } \label{fig_ratios} 
\end{figure}

We computed X/Fe ratio profiles for Si, S, Ar, and Ca (gray shaded regions in Figure~\ref{fig_ratios}). Considering the uncertainties, we have found Si/Fe and S/Fe ratios close to solar values for all distances. On the other hand, the Ar/Fe and Ca/Fe ratios are higher than solar. The former one, in particular, has very high values ($>3$) for large distances, although uncertainties are considerable. This is the first time such ratios are measured within the Ophiuchus cluster. 

We used the {\tt SNeRatio} python code developed by \citep{erd21} to estimate the contribution from different SN yield models to the abundance ratios obtained. For a given set of ICM abundances, the code fits it with a combination of multiple progenitor yield models to compute the relative contribution that better fits the observed data. We included models from \citet{nom13} with initial metallicity values of Z $=$ 0.0, 0.001, 0.004, 0.008, 0.02, 0.05 for the SNcc yields, integrated with Salpeter IMF over the mass range of 10-70 M$_\odot$.  We consider a set of 3D SNIa models near Chandrasekhar-mass, including pure deflagration from \citet{fin14} and delayed detonation from \citet{sei13} as possible explosion mechanisms. For all distances, we assumed the same SNe model. Thus, we determined the best linear combination of SNia and SNcc models that better fit the abundance ratio profiles by minimizing the sum of their $\chi^{2}$ values. Such models have been used in previous enrichment studies \citep{mer17,sim19,mer20,gat23b}. However, we cannot ignore that current yield calculations are prone to significant uncertainties. 

\begin{figure}    
\centering
\includegraphics[width=0.45\textwidth]{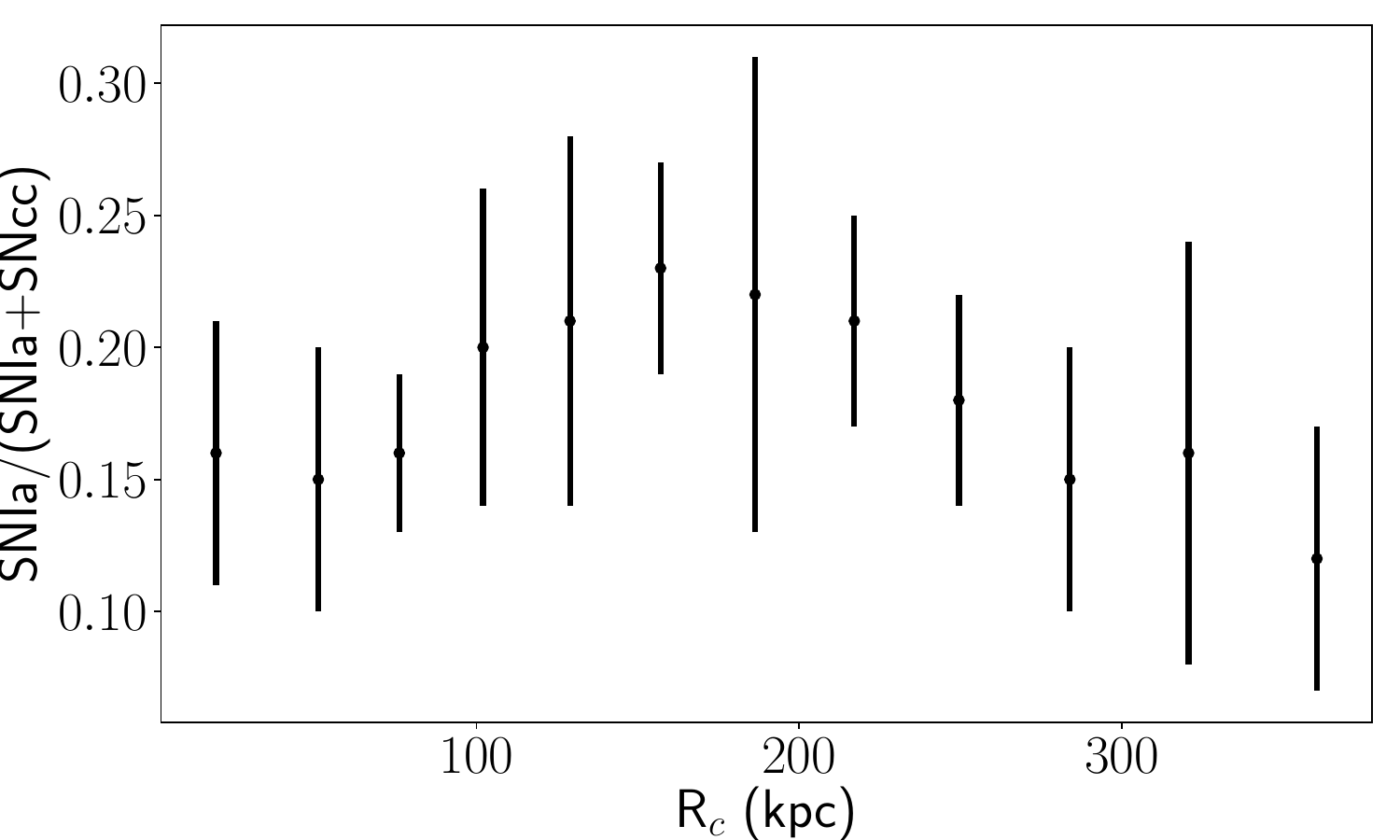} 
\caption{
SNIa contribution to the total chemical enrichment as function of the distance. The model includes Si, S, Ar, Ca and Fe abundances.
} \label{fig_snia_distribution} 
\end{figure}

We found that the best-fit model corresponds to a delayed detonation 3D N100 model for SNIa \citep[see Table~1 in][]{sei13} and SNcc model with initial metallicity Z$=0.01$. The model is shown in Figure~\ref{fig_ratios} (red-shaded regions). The contribution from SNIa to the total enrichment by this set of models is shown in Table~\ref{tab_snr_contribution} and Figure~\ref{fig_snia_distribution}. The Ca/Fe and Ar/Fe ratios are not reproduced satisfactorily at large distances. We have found a small SNIa contribution to the total SN model for almost all radii ($10-30\%$). A linear fit gives almost zero slope (1.98$\times$10$^{-5}$ with $\chi^{2}=1.03$) and a constant value of $0.18\pm 0.02$. Such a consistent radial profile may suggest that the SNIcc and SNIa components have occurred at similar epochs. The most established assumption to explain such metal distribution is that most metals released into the ICM happened before the cluster assembly \citep{mer22}. Nevertheless, it is important to highlight that the absence of prominent emission characteristics originating from light $\alpha$-elements (i.e., O, Ne, Mg) could potentially result in an overestimation of the contribution from SNcc, as it is the primary source of production of these elements. Furthermore, list Table~\ref{tab_snr_models} shows a list of SN yield models for which good fits are obtained (i.e., $\chi^{2}<1.2$). All the best-fit models correspond to delayed detonation 3D models from \citet{sei13}. Given the uncertainties in the SN models, such models cannot be dismissed entirely. Finally, a linear combination of SN Ia and SNcc does not reproduce the observed ratios may be partly due to the assumption that a few classes of supernovae dominate the chemical enrichment \citep[see ][]{sim19}.

\begin{table}
\caption{\label{tab_snr_contribution}SNIa contributions to the total chemical enrichment. }
\centering
\begin{tabular}{ccccccc}
\\
Radius  & SNIa & Radius & SNIa \\
 (kpc) & & (kpc)   & \\ 
\hline 
19.4 & $16\pm 5 \% $ &186.4& $22\pm 9 \% $\\ 
51.0 & $15\pm 5\% $ &217.1& $21\pm 4 \% $\\ 
76.1 &$16\pm 3 \% $  &249.5&$18\pm 4 \% $ \\
101.1&$20\pm 6 \% $  &283.9&$15\pm 5 \% $ \\ 
129.0 &$21\pm 7 \% $ &320.7& $16\pm 8 \% $\\ 
157.1 &$23\pm 4 \% $ &360.4& $12\pm 5 \% $\\ 
\\ 
 \hline
\end{tabular}
\end{table}

\begin{table}
\caption{\label{tab_snr_models} $\chi^{2}$ fit-values obtained for a sample of SN models. }
\centering
\begin{tabular}{ccc}
\\
\multicolumn{2}{c}{Model} & $\chi^{2}$ \\  
SNIa & SNcc &  \\ 
\hline 
N1 - \citet{sei13}   & Z=0.008 & $1.11$\\  
N1 - \citet{sei13}   & Z=0.02 & $1.11$\\  
N10 - \citet{sei13} & Z=0.008 & $1.13$\\  
N10 - \citet{sei13}  & Z=0.02 & $1.13$\\ 
N3 - \citet{sei13}   & Z=0.008 & $1.13$\\
N3 - \citet{sei13}   & Z=0.02 & $1.15$\\
N5 - \citet{sei13}  & Z=0.02 & $1.17$\\  
 \hline
\end{tabular}
\end{table}

\section{Conclusions and summary}\label{sec_con} 
We have analyzed ICM radial profiles within the Ophiuchus galaxy cluster using {\it XMM-Newton} EPIC-pn observations to study the distribution of physical parameters such as hydrogen column density, temperature, and distribution Si, S, Ar, Ca, and Fe. This work expands upon the outcomes of \citet{gat23a} by including the soft energy band  $<4$~keV. In this Section, we briefly highlight our results

\begin{enumerate}
\item We model the ICM emission using a log-normal temperature distribution model namely {\tt lognorm}. Such a model performs a better fit of the data than a single or double-temperature model. 
\item We report a hydrogen column density ($N_{\rm H}$) profile from the cluster core to the outermost region. Furthermore, we found higher values than those obtained from radio measurements, most likely due to the presence of dust and molecular gas in the line of sight.
\item We found that the temperature steeply increases for distances $< 100$~kpc and then continues increasing at a slower rate, with a peak value of $11.67\pm 0.95$~keV.
\item The best-fit velocities obtained are in good agreement with the previous work by \citet{gat23a}, even when here we included the soft energy band. This illustrates that the method for measuring velocities is reliable. We found that most of the velocities show no deviation from the cluster velocity.
\item We obtained radial profiles for Si, S, Ar, Ca, and Fe. Because the system is highly absorbed, we have obtained only upper limits for O, Ne, Mg, and Ni. The S, Si, and Fe radial distributions display a decreasing behavior as a function of the distance. At the same time, we have not found changes in the abundance profiles with statistical significance $>2\sigma$ for the other elements. 
\item We compute X/Fe abundance ratio profiles for Si, S, Ar, and Ca. We model these profiles with a linear combination of SNcc and SNIa. We found that the best-fit model corresponds to a delayed detonation 3D  model for SNIa and an initial metallicity Z$=0.01$ for SNcc. While this model roughly reproduces the obtained Si/Fe and S/Fe profiles, the Ar/Fe and Ca/Fe profiles are not well reproduced. 
\item We found a flat radial distribution of SNIa ratio over the total cluster enrichment $10-30\%$ for all radii. However, the absence of light $\alpha$-elements abundances may lead to over-estimation of the SNcc contribution.

\end{enumerate} 

An extensive analysis of the 2D abundance spatial distribution will be presented in a forthcoming paper.

\section{Acknowledgements} 
This work was supported by the Deutsche Zentrum f\"ur Luft- und Raumfahrt (DLR) under the Verbundforschung programme (Messung von Schwapp-, Verschmelzungs- und R\"uckkopplungs-geschwindigkeiten in Galaxienhaufen). This work is based on observations obtained with XMM-Newton, an ESA science mission with instruments and contributions directly funded by ESA Member States and NASA. This research was carried out on the High Performance Computing resources of the cobra cluster at the Max Planck Computing and Data Facility (MPCDF) in Garching operated by the Max Planck Society (MPG).
\subsection*{Data availability}
The observations analyzed in this article are available in the {\it XMM-Newton} Science Archive (XSA\footnote{\url{http://xmm.esac.esa.int/xsa/}}).

\bibliographystyle{mnras}

\begin{thebibliography}{}
\makeatletter
\relax
\def\mn@urlcharsother{\let\do\@makeother \do\$\do\&\do\#\do\^\do\_\do\%\do\~}
\def\mn@doi{\begingroup\mn@urlcharsother \@ifnextchar [ {\mn@doi@}
  {\mn@doi@[]}}
\def\mn@doi@[#1]#2{\def\@tempa{#1}\ifx\@tempa\@empty \href
  {http://dx.doi.org/#2} {doi:#2}\else \href {http://dx.doi.org/#2} {#1}\fi
  \endgroup}
\def\mn@eprint#1#2{\mn@eprint@#1:#2::\@nil}
\def\mn@eprint@arXiv#1{\href {http://arxiv.org/abs/#1} {{\tt arXiv:#1}}}
\def\mn@eprint@dblp#1{\href {http://dblp.uni-trier.de/rec/bibtex/#1.xml}
  {dblp:#1}}
\def\mn@eprint@#1:#2:#3:#4\@nil{\def\@tempa {#1}\def\@tempb {#2}\def\@tempc
  {#3}\ifx \@tempc \@empty \let \@tempc \@tempb \let \@tempb \@tempa \fi \ifx
  \@tempb \@empty \def\@tempb {arXiv}\fi \@ifundefined
  {mn@eprint@\@tempb}{\@tempb:\@tempc}{\expandafter \expandafter \csname
  mn@eprint@\@tempb\endcsname \expandafter{\@tempc}}}

\bibitem[\protect\citeauthoryear{{Bohlin}, {Savage}  \& {Drake}}{{Bohlin}
  et~al.}{1978}]{boh78}
{Bohlin} R.~C.,  {Savage} B.~D.,   {Drake} J.~F.,  1978, \mn@doi [\apj]
  {10.1086/156357}, \href
  {https://ui.adsabs.harvard.edu/abs/1978ApJ...224..132B} {224, 132}

\bibitem[\protect\citeauthoryear{{Cash}}{{Cash}}{1979}]{cas79}
{Cash} W.,  1979, \mn@doi [\apj] {10.1086/156922}, \href
  {http://adsabs.harvard.edu/abs/1979ApJ...228..939C} {228, 939}

\bibitem[\protect\citeauthoryear{{Churazov}, {Forman}, {Jones}  \&
  {B{\"o}hringer}}{{Churazov} et~al.}{2003}]{chu03}
{Churazov} E.,  {Forman} W.,  {Jones} C.,   {B{\"o}hringer} H.,  2003, \mn@doi
  [\apj] {10.1086/374923}, \href
  {https://ui.adsabs.harvard.edu/abs/2003ApJ...590..225C} {590, 225}

\bibitem[\protect\citeauthoryear{{Das}, {Mathur}, {Gupta}  \& {Krongold}}{{Das}
  et~al.}{2021}]{das21}
{Das} S.,  {Mathur} S.,  {Gupta} A.,   {Krongold} Y.,  2021, \mn@doi [\apj]
  {10.3847/1538-4357/ac0e8e}, \href
  {https://ui.adsabs.harvard.edu/abs/2021ApJ...918...83D} {918, 83}

\bibitem[\protect\citeauthoryear{{De Grandi} \& {Molendi}}{{De Grandi} \&
  {Molendi}}{2001}]{deg01}
{De Grandi} S.,  {Molendi} S.,  2001, \mn@doi [\apj] {10.1086/320098}, \href
  {https://ui.adsabs.harvard.edu/abs/2001ApJ...551..153D} {551, 153}

\bibitem[\protect\citeauthoryear{{Durret}, {Wakamatsu}, {Nagayama}, {Adami}  \&
  {Biviano}}{{Durret} et~al.}{2015}]{dur15}
{Durret} F.,  {Wakamatsu} K.,  {Nagayama} T.,  {Adami} C.,   {Biviano} A.,
  2015, \mn@doi [\aap] {10.1051/0004-6361/201526531}, \href
  {https://ui.adsabs.harvard.edu/abs/2015A&A...583A.124D} {583, A124}

\bibitem[\protect\citeauthoryear{{Eckert}, {Ettori}, {Pointecouteau},
  {Molendi}, {Paltani}  \& {Tchernin}}{{Eckert} et~al.}{2017}]{eck17}
{Eckert} D.,  {Ettori} S.,  {Pointecouteau} E.,  {Molendi} S.,  {Paltani} S.,
  {Tchernin} C.,  2017, \mn@doi [Astronomische Nachrichten]
  {10.1002/asna.201713345}, \href
  {https://ui.adsabs.harvard.edu/abs/2017AN....338..293E} {338, 293}

\bibitem[\protect\citeauthoryear{{Edge}, {Stewart}, {Fabian}  \&
  {Arnaud}}{{Edge} et~al.}{1990}]{edg90}
{Edge} A.~C.,  {Stewart} G.~C.,  {Fabian} A.~C.,   {Arnaud} K.~A.,  1990,
  \mnras, \href {https://ui.adsabs.harvard.edu/abs/1990MNRAS.245..559E} {245,
  559}

\bibitem[\protect\citeauthoryear{{Erdim}, {Ezer}, {{\"U}nver}, {Hazar}  \&
  {Hudaverdi}}{{Erdim} et~al.}{2021}]{erd21}
{Erdim} M.~K.,  {Ezer} C.,  {{\"U}nver} O.,  {Hazar} F.,   {Hudaverdi} M.,
  2021, \mn@doi [\mnras] {10.1093/mnras/stab2730}, \href
  {https://ui.adsabs.harvard.edu/abs/2021MNRAS.508.3337E} {508, 3337}

\bibitem[\protect\citeauthoryear{{Fabian}, {Ferland}, {Sanders}, {McNamara},
  {Pinto}  \& {Walker}}{{Fabian} et~al.}{2022}]{fab22b}
{Fabian} A.~C.,  {Ferland} G.~J.,  {Sanders} J.~S.,  {McNamara} B.~R.,  {Pinto}
  C.,   {Walker} S.~A.,  2022, \mn@doi [\mnras] {10.1093/mnras/stac2003}, \href
  {https://ui.adsabs.harvard.edu/abs/2022MNRAS.515.3336F} {515, 3336}

\bibitem[\protect\citeauthoryear{{Fabian}, {Sanders}, {Ferland}, {McNamara},
  {Pinto}  \& {Walker}}{{Fabian} et~al.}{2023a}]{fab23a}
{Fabian} A.~C.,  {Sanders} J.~S.,  {Ferland} G.~J.,  {McNamara} B.~R.,  {Pinto}
  C.,   {Walker} S.~A.,  2023a, \mn@doi [\mnras] {10.1093/mnras/stad1870},
  \href {https://ui.adsabs.harvard.edu/abs/2023MNRAS.tmp.1807F} {}

\bibitem[\protect\citeauthoryear{{Fabian}, {Sanders}, {Ferland}, {McNamara},
  {Pinto}  \& {Walker}}{{Fabian} et~al.}{2023b}]{fab23b}
{Fabian} A.~C.,  {Sanders} J.~S.,  {Ferland} G.~J.,  {McNamara} B.~R.,  {Pinto}
  C.,   {Walker} S.~A.,  2023b, \mn@doi [\mnras] {10.1093/mnras/stad507}, \href
  {https://ui.adsabs.harvard.edu/abs/2023MNRAS.521.1794F} {521, 1794}

\bibitem[\protect\citeauthoryear{{Fink} et~al.,}{{Fink} et~al.}{2014}]{fin14}
{Fink} M.,  et~al., 2014, \mn@doi [\mnras] {10.1093/mnras/stt2315}, \href
  {https://ui.adsabs.harvard.edu/abs/2014MNRAS.438.1762F} {438, 1762}

\bibitem[\protect\citeauthoryear{{Frank}, {Peterson}, {Andersson}, {Fabian}  \&
  {Sanders}}{{Frank} et~al.}{2013}]{fra13}
{Frank} K.~A.,  {Peterson} J.~R.,  {Andersson} K.,  {Fabian} A.~C.,   {Sanders}
  J.~S.,  2013, \mn@doi [\apj] {10.1088/0004-637X/764/1/46}, \href
  {https://ui.adsabs.harvard.edu/abs/2013ApJ...764...46F} {764, 46}

\bibitem[\protect\citeauthoryear{{Fujita} et~al.,}{{Fujita}
  et~al.}{2008}]{fuj08}
{Fujita} Y.,  et~al., 2008, \mn@doi [\pasj] {10.1093/pasj/60.5.1133}, \href
  {https://ui.adsabs.harvard.edu/abs/2008PASJ...60.1133F} {60, 1133}

\bibitem[\protect\citeauthoryear{{Fukushima}, {Kobayashi}  \&
  {Matsushita}}{{Fukushima} et~al.}{2022}]{fuk22}
{Fukushima} K.,  {Kobayashi} S.~B.,   {Matsushita} K.,  2022, \mn@doi [\mnras]
  {10.1093/mnras/stac1590}, \href
  {https://ui.adsabs.harvard.edu/abs/2022MNRAS.514.4222F} {514, 4222}

\bibitem[\protect\citeauthoryear{{Gatuzz}, {Sanders}, {Dennerl}, {Pinto},
  {Fabian}, {Tamura}, {Walker}  \& {ZuHone}}{{Gatuzz} et~al.}{2022a}]{gat22a}
{Gatuzz} E.,  {Sanders} J.~S.,  {Dennerl} K.,  {Pinto} C.,  {Fabian} A.~C.,
  {Tamura} T.,  {Walker} S.~A.,   {ZuHone} J.,  2022a, \mn@doi [\mnras]
  {10.1093/mnras/stab2661}, \href
  {https://ui.adsabs.harvard.edu/abs/2022MNRAS.511.4511G} {511, 4511}

\bibitem[\protect\citeauthoryear{{Gatuzz} et~al.,}{{Gatuzz}
  et~al.}{2022b}]{gat22b}
{Gatuzz} E.,  et~al., 2022b, \mn@doi [\mnras] {10.1093/mnras/stac846}, \href
  {https://ui.adsabs.harvard.edu/abs/2022MNRAS.513.1932G} {513, 1932}

\bibitem[\protect\citeauthoryear{{Gatuzz} et~al.,}{{Gatuzz}
  et~al.}{2023a}]{gat23a}
{Gatuzz} E.,  et~al., 2023a, \mn@doi [arXiv e-prints]
  {10.48550/arXiv.2303.17556}, \href
  {https://ui.adsabs.harvard.edu/abs/2023arXiv230317556G} {p. arXiv:2303.17556}

\bibitem[\protect\citeauthoryear{{Gatuzz} et~al.,}{{Gatuzz}
  et~al.}{2023b}]{gat23b}
{Gatuzz} E.,  et~al., 2023b, \mn@doi [\mnras] {10.1093/mnras/stad447}, \href
  {https://ui.adsabs.harvard.edu/abs/2023MNRAS.520.4793G} {520, 4793}

\bibitem[\protect\citeauthoryear{{Ghizzardi} et~al.,}{{Ghizzardi}
  et~al.}{2021}]{sim21}
{Ghizzardi} S.,  et~al., 2021, \mn@doi [\aap] {10.1051/0004-6361/202038501},
  \href {https://ui.adsabs.harvard.edu/abs/2021A&A...646A..92G} {646, A92}

\bibitem[\protect\citeauthoryear{{Giacintucci}, {Markevitch},
  {Johnston-Hollitt}, {Wik}, {Wang}  \& {Clarke}}{{Giacintucci}
  et~al.}{2020}]{gia20}
{Giacintucci} S.,  {Markevitch} M.,  {Johnston-Hollitt} M.,  {Wik} D.~R.,
  {Wang} Q.~H.~S.,   {Clarke} T.~E.,  2020, \mn@doi [\apj]
  {10.3847/1538-4357/ab6a9d}, \href
  {https://ui.adsabs.harvard.edu/abs/2020ApJ...891....1G} {891, 1}

\bibitem[\protect\citeauthoryear{{Gupta}, {Mathur}, {Kingsbury}, {Das}  \&
  {Krongold}}{{Gupta} et~al.}{2023}]{gup23}
{Gupta} A.,  {Mathur} S.,  {Kingsbury} J.,  {Das} S.,   {Krongold} Y.,  2023,
  \mn@doi [Nature Astronomy] {10.1038/s41550-023-01963-5}, \href
  {https://ui.adsabs.harvard.edu/abs/2023NatAs.tmp...91G} {}

\bibitem[\protect\citeauthoryear{{Hitomi Collaboration} et~al.,}{{Hitomi
  Collaboration} et~al.}{2018}]{hit18}
{Hitomi Collaboration} et~al., 2018, \mn@doi [\pasj] {10.1093/pasj/psx156},
  \href {https://ui.adsabs.harvard.edu/abs/2018PASJ...70...12H} {70, 12}

\bibitem[\protect\citeauthoryear{{Kalberla}, {Burton}, {Hartmann}, {Arnal},
  {Bajaja}, {Morras}  \& {P{\"o}ppel}}{{Kalberla} et~al.}{2005}]{kal05}
{Kalberla} P.~M.~W.,  {Burton} W.~B.,  {Hartmann} D.,  {Arnal} E.~M.,  {Bajaja}
  E.,  {Morras} R.,   {P{\"o}ppel} W.~G.~L.,  2005, \mn@doi [\aap]
  {10.1051/0004-6361:20041864}, \href
  {http://adsabs.harvard.edu/abs/2005A%26A...440..775K} {440, 775}

\bibitem[\protect\citeauthoryear{{Liu}, {Zhai}  \& {Tozzi}}{{Liu}
  et~al.}{2019}]{liu19}
{Liu} A.,  {Zhai} M.,   {Tozzi} P.,  2019, \mn@doi [\mnras]
  {10.1093/mnras/stz533}, \href
  {https://ui.adsabs.harvard.edu/abs/2019MNRAS.485.1651L} {485, 1651}

\bibitem[\protect\citeauthoryear{{Liu}, {Tozzi}, {Ettori}, {De Grandi},
  {Gastaldello}, {Rosati}  \& {Norman}}{{Liu} et~al.}{2020}]{liu20}
{Liu} A.,  {Tozzi} P.,  {Ettori} S.,  {De Grandi} S.,  {Gastaldello} F.,
  {Rosati} P.,   {Norman} C.,  2020, \mn@doi [\aap]
  {10.1051/0004-6361/202037506}, \href
  {https://ui.adsabs.harvard.edu/abs/2020A&A...637A..58L} {637, A58}

\bibitem[\protect\citeauthoryear{{Lodders} \& {Palme}}{{Lodders} \&
  {Palme}}{2009}]{lod09}
{Lodders} K.,  {Palme} H.,  2009, Meteoritics and Planetary Science Supplement,
  \href {https://ui.adsabs.harvard.edu/abs/2009M&PSA..72.5154L} {72, 5154}

\bibitem[\protect\citeauthoryear{{Lovisari} \& {Reiprich}}{{Lovisari} \&
  {Reiprich}}{2019}]{lov19}
{Lovisari} L.,  {Reiprich} T.~H.,  2019, \mn@doi [\mnras]
  {10.1093/mnras/sty3130}, \href
  {https://ui.adsabs.harvard.edu/abs/2019MNRAS.483..540L} {483, 540}

\bibitem[\protect\citeauthoryear{{Matsushita}}{{Matsushita}}{2011}]{mat11}
{Matsushita} K.,  2011, \mn@doi [\aap] {10.1051/0004-6361/200913432}, \href
  {https://ui.adsabs.harvard.edu/abs/2011A&A...527A.134M} {527, A134}

\bibitem[\protect\citeauthoryear{{Mernier} \& {Biffi}}{{Mernier} \&
  {Biffi}}{2022}]{mer22}
{Mernier} F.,  {Biffi} V.,  2022, in , Handbook of X-ray and Gamma-ray
  Astrophysics. Edited by Cosimo Bambi and Andrea Santangelo.
p.~12, \mn@doi{10.1007/978-981-16-4544-0_123-1}

\bibitem[\protect\citeauthoryear{{Mernier}, {de Plaa}, {Lovisari}, {Pinto},
  {Zhang}, {Kaastra}, {Werner}  \& {Simionescu}}{{Mernier}
  et~al.}{2015}]{mer15}
{Mernier} F.,  {de Plaa} J.,  {Lovisari} L.,  {Pinto} C.,  {Zhang} Y.~Y.,
  {Kaastra} J.~S.,  {Werner} N.,   {Simionescu} A.,  2015, \mn@doi [\aap]
  {10.1051/0004-6361/201425282}, \href
  {https://ui.adsabs.harvard.edu/abs/2015A&A...575A..37M} {575, A37}

\bibitem[\protect\citeauthoryear{{Mernier}, {de Plaa}, {Pinto}, {Kaastra},
  {Kosec}, {Zhang}, {Mao}  \& {Werner}}{{Mernier} et~al.}{2016}]{mer16}
{Mernier} F.,  {de Plaa} J.,  {Pinto} C.,  {Kaastra} J.~S.,  {Kosec} P.,
  {Zhang} Y.~Y.,  {Mao} J.,   {Werner} N.,  2016, \mn@doi [\aap]
  {10.1051/0004-6361/201527824}, \href
  {https://ui.adsabs.harvard.edu/abs/2016A&A...592A.157M} {592, A157}

\bibitem[\protect\citeauthoryear{{Mernier} et~al.,}{{Mernier}
  et~al.}{2017}]{mer17}
{Mernier} F.,  et~al., 2017, \mn@doi [\aap] {10.1051/0004-6361/201630075},
  \href {https://ui.adsabs.harvard.edu/abs/2017A&A...603A..80M} {603, A80}

\bibitem[\protect\citeauthoryear{{Mernier} et~al.,}{{Mernier}
  et~al.}{2018}]{mer18}
{Mernier} F.,  et~al., 2018, \mn@doi [\ssr] {10.1007/s11214-018-0565-7}, \href
  {https://ui.adsabs.harvard.edu/abs/2018SSRv..214..129M} {214, 129}

\bibitem[\protect\citeauthoryear{{Mernier} et~al.,}{{Mernier}
  et~al.}{2020}]{mer20}
{Mernier} F.,  et~al., 2020, \mn@doi [\aap] {10.1051/0004-6361/202038638},
  \href {https://ui.adsabs.harvard.edu/abs/2020A&A...642A..90M} {642, A90}

\bibitem[\protect\citeauthoryear{{Million}, {Allen}, {Werner}  \&
  {Taylor}}{{Million} et~al.}{2010}]{mil10}
{Million} E.~T.,  {Allen} S.~W.,  {Werner} N.,   {Taylor} G.~B.,  2010, \mn@doi
  [\mnras] {10.1111/j.1365-2966.2010.16596.x}, \href
  {https://ui.adsabs.harvard.edu/abs/2010MNRAS.405.1624M} {405, 1624}

\bibitem[\protect\citeauthoryear{{Nevalainen}, {Eckert}, {Kaastra}, {Bonamente}
   \& {Kettula}}{{Nevalainen} et~al.}{2009}]{nev09}
{Nevalainen} J.,  {Eckert} D.,  {Kaastra} J.,  {Bonamente} M.,   {Kettula} K.,
  2009, \mn@doi [\aap] {10.1051/0004-6361/200912542}, \href
  {https://ui.adsabs.harvard.edu/abs/2009A&A...508.1161N} {508, 1161}

\bibitem[\protect\citeauthoryear{{Nomoto}, {Kobayashi}  \& {Tominaga}}{{Nomoto}
  et~al.}{2013}]{nom13}
{Nomoto} K.,  {Kobayashi} C.,   {Tominaga} N.,  2013, \mn@doi [\araa]
  {10.1146/annurev-astro-082812-140956}, \href
  {https://ui.adsabs.harvard.edu/abs/2013ARA&A..51..457N} {51, 457}

\bibitem[\protect\citeauthoryear{{Panagoulia}, {Fabian}  \&
  {Sanders}}{{Panagoulia} et~al.}{2013}]{pan13}
{Panagoulia} E.~K.,  {Fabian} A.~C.,   {Sanders} J.~S.,  2013, \mn@doi [\mnras]
  {10.1093/mnras/stt969}, \href
  {https://ui.adsabs.harvard.edu/abs/2013MNRAS.433.3290P} {433, 3290}

\bibitem[\protect\citeauthoryear{{Panagoulia}, {Sanders}  \&
  {Fabian}}{{Panagoulia} et~al.}{2015}]{pan15}
{Panagoulia} E.~K.,  {Sanders} J.~S.,   {Fabian} A.~C.,  2015, \mn@doi [\mnras]
  {10.1093/mnras/stu2469}, \href
  {https://ui.adsabs.harvard.edu/abs/2015MNRAS.447..417P} {447, 417}

\bibitem[\protect\citeauthoryear{{P{\'e}rez-Torres}, {Zandanel}, {Guerrero},
  {Pal}, {Profumo}, {Prada}  \& {Panessa}}{{P{\'e}rez-Torres}
  et~al.}{2009}]{per09}
{P{\'e}rez-Torres} M.~A.,  {Zandanel} F.,  {Guerrero} M.~A.,  {Pal} S.,
  {Profumo} S.,  {Prada} F.,   {Panessa} F.,  2009, \mn@doi [\mnras]
  {10.1111/j.1365-2966.2009.14883.x}, \href
  {https://ui.adsabs.harvard.edu/abs/2009MNRAS.396.2237P} {396, 2237}

\bibitem[\protect\citeauthoryear{{Planelles}, {Borgani}, {Fabjan}, {Killedar},
  {Murante}, {Granato}, {Ragone-Figueroa}  \& {Dolag}}{{Planelles}
  et~al.}{2014}]{pla14}
{Planelles} S.,  {Borgani} S.,  {Fabjan} D.,  {Killedar} M.,  {Murante} G.,
  {Granato} G.~L.,  {Ragone-Figueroa} C.,   {Dolag} K.,  2014, \mn@doi [\mnras]
  {10.1093/mnras/stt2141}, \href
  {https://ui.adsabs.harvard.edu/abs/2014MNRAS.438..195P} {438, 195}

\bibitem[\protect\citeauthoryear{{Sanders} \& {Fabian}}{{Sanders} \&
  {Fabian}}{2002}]{san02}
{Sanders} J.~S.,  {Fabian} A.~C.,  2002, \mn@doi [\mnras]
  {10.1046/j.1365-8711.2002.05211.x}, \href
  {https://ui.adsabs.harvard.edu/abs/2002MNRAS.331..273S} {331, 273}

\bibitem[\protect\citeauthoryear{{Sanders} et~al.,}{{Sanders}
  et~al.}{2016}]{san16}
{Sanders} J.~S.,  et~al., 2016, \mn@doi [\mnras] {10.1093/mnras/stv2972}, \href
  {https://ui.adsabs.harvard.edu/abs/2016MNRAS.457...82S} {457, 82}

\bibitem[\protect\citeauthoryear{{Sanders} et~al.,}{{Sanders}
  et~al.}{2020}]{san20}
{Sanders} J.~S.,  et~al., 2020, \mn@doi [\aap] {10.1051/0004-6361/201936468},
  \href {https://ui.adsabs.harvard.edu/abs/2020A&A...633A..42S} {633, A42}

\bibitem[\protect\citeauthoryear{{Schellenberger}, {Reiprich}, {Lovisari},
  {Nevalainen}  \& {David}}{{Schellenberger} et~al.}{2015}]{sch15}
{Schellenberger} G.,  {Reiprich} T.~H.,  {Lovisari} L.,  {Nevalainen} J.,
  {David} L.,  2015, \mn@doi [\aap] {10.1051/0004-6361/201424085}, \href
  {https://ui.adsabs.harvard.edu/abs/2015A&A...575A..30S} {575, A30}

\bibitem[\protect\citeauthoryear{{Schlegel}, {Finkbeiner}  \&
  {Davis}}{{Schlegel} et~al.}{1998}]{sch98}
{Schlegel} D.~J.,  {Finkbeiner} D.~P.,   {Davis} M.,  1998, \mn@doi [\apj]
  {10.1086/305772}, \href
  {https://ui.adsabs.harvard.edu/abs/1998ApJ...500..525S} {500, 525}

\bibitem[\protect\citeauthoryear{{Seitenzahl} et~al.,}{{Seitenzahl}
  et~al.}{2013}]{sei13}
{Seitenzahl} I.~R.,  et~al., 2013, \mn@doi [\mnras] {10.1093/mnras/sts402},
  \href {https://ui.adsabs.harvard.edu/abs/2013MNRAS.429.1156S} {429, 1156}

\bibitem[\protect\citeauthoryear{{Simionescu}, {Werner}, {Urban}, {Allen},
  {Ichinohe}  \& {Zhuravleva}}{{Simionescu} et~al.}{2015}]{sim15}
{Simionescu} A.,  {Werner} N.,  {Urban} O.,  {Allen} S.~W.,  {Ichinohe} Y.,
  {Zhuravleva} I.,  2015, \mn@doi [\apjl] {10.1088/2041-8205/811/2/L25}, \href
  {https://ui.adsabs.harvard.edu/abs/2015ApJ...811L..25S} {811, L25}

\bibitem[\protect\citeauthoryear{{Simionescu} et~al.,}{{Simionescu}
  et~al.}{2019}]{sim19}
{Simionescu} A.,  et~al., 2019, \mn@doi [\mnras] {10.1093/mnras/sty3220}, \href
  {https://ui.adsabs.harvard.edu/abs/2019MNRAS.483.1701S} {483, 1701}

\bibitem[\protect\citeauthoryear{{Str{\"u}der} et~al.,}{{Str{\"u}der}
  et~al.}{2001}]{str01}
{Str{\"u}der} L.,  et~al., 2001, \mn@doi [\aap] {10.1051/0004-6361:20000066},
  \href {http://adsabs.harvard.edu/abs/2001A%26A...365L..18S} {365, L18}

\bibitem[\protect\citeauthoryear{{Sun}, {Murray}, {Markevitch}  \&
  {Vikhlinin}}{{Sun} et~al.}{2002}]{sun02}
{Sun} M.,  {Murray} S.~S.,  {Markevitch} M.,   {Vikhlinin} A.,  2002, \mn@doi
  [\apj] {10.1086/324721}, \href
  {https://ui.adsabs.harvard.edu/abs/2002ApJ...565..867S} {565, 867}

\bibitem[\protect\citeauthoryear{{Urban}, {Werner}, {Allen}, {Simionescu}  \&
  {Mantz}}{{Urban} et~al.}{2017}]{urb17}
{Urban} O.,  {Werner} N.,  {Allen} S.~W.,  {Simionescu} A.,   {Mantz} A.,
  2017, \mn@doi [\mnras] {10.1093/mnras/stx1542}, \href
  {https://ui.adsabs.harvard.edu/abs/2017MNRAS.470.4583U} {470, 4583}

\bibitem[\protect\citeauthoryear{{Urdampilleta}, {Mernier}, {Kaastra},
  {Simionescu}, {de Plaa}, {Kara}  \& {Ercan}}{{Urdampilleta}
  et~al.}{2019}]{urd19}
{Urdampilleta} I.,  {Mernier} F.,  {Kaastra} J.~S.,  {Simionescu} A.,  {de
  Plaa} J.,  {Kara} S.,   {Ercan} E.~N.,  2019, \mn@doi [\aap]
  {10.1051/0004-6361/201935452}, \href
  {https://ui.adsabs.harvard.edu/abs/2019A&A...629A..31U} {629, A31}

\bibitem[\protect\citeauthoryear{{Walker} \& {Lau}}{{Walker} \&
  {Lau}}{2022}]{wal22}
{Walker} S.,  {Lau} E.,  2022, in , Handbook of X-ray and Gamma-ray
  Astrophysics.
p.~13, \mn@doi{10.1007/978-981-16-4544-0_120-1}

\bibitem[\protect\citeauthoryear{{Werner}, {Durret}, {Ohashi}, {Schindler}  \&
  {Wiersma}}{{Werner} et~al.}{2008}]{wer08}
{Werner} N.,  {Durret} F.,  {Ohashi} T.,  {Schindler} S.,   {Wiersma} R.~P.~C.,
   2008, \mn@doi [\ssr] {10.1007/s11214-008-9320-9}, \href
  {https://ui.adsabs.harvard.edu/abs/2008SSRv..134..337W} {134, 337}

\bibitem[\protect\citeauthoryear{{Werner}, {Urban}, {Simionescu}  \&
  {Allen}}{{Werner} et~al.}{2013}]{wer13}
{Werner} N.,  {Urban} O.,  {Simionescu} A.,   {Allen} S.~W.,  2013, \mn@doi
  [\nat] {10.1038/nature12646}, \href
  {https://ui.adsabs.harvard.edu/abs/2013Natur.502..656W} {502, 656}

\bibitem[\protect\citeauthoryear{{Werner} et~al.,}{{Werner}
  et~al.}{2016a}]{wer16b}
{Werner} N.,  et~al., 2016a, \mn@doi [\mnras] {10.1093/mnras/stw1171}, \href
  {https://ui.adsabs.harvard.edu/abs/2016MNRAS.460.2752W} {460, 2752}

\bibitem[\protect\citeauthoryear{{Werner} et~al.,}{{Werner}
  et~al.}{2016b}]{wer16}
{Werner} N.,  et~al., 2016b, \mn@doi [\mnras] {10.1093/mnras/stw1171}, \href
  {https://ui.adsabs.harvard.edu/abs/2016MNRAS.460.2752W} {460, 2752}

\bibitem[\protect\citeauthoryear{{Yoshino} et~al.,}{{Yoshino}
  et~al.}{2009}]{yos09}
{Yoshino} T.,  et~al., 2009, \mn@doi [\pasj] {10.1093/pasj/61.4.805}, \href
  {https://ui.adsabs.harvard.edu/abs/2009PASJ...61..805Y} {61, 805}

\makeatother
\end{thebibliography}
 \newcommand{\noop}[1]{}

\end{document}